\documentclass[a4paper,10pt]{article}
\usepackage[utf8]{inputenc}
\usepackage{graphicx}
\usepackage{hyperref}
\usepackage{color}
\usepackage[left=2cm,right=2cm]{geometry}
\usepackage{amsmath,amsfonts,amssymb}
\usepackage{cite}
\usepackage{csquotes}
\usepackage{authblk}

\makeatother

\title{Spacetime development of in-medium hadronization: Scenario for leading hadrons}
\author{B. Guiot\footnote{benjamin.guiot@usm.cl} \, and B. Z. Kopeliovich\footnote{boris.kopeliovich@usm.cl}}
\affil[]{\small{Departamento de F\'isica, Universidad T\'ecnica Federico Santa Mar\'ia; Casilla 110-V, Valparaiso, Chile}}
\date{}

\begin{document}

\maketitle

\begin{abstract}
We present a perturbative QCD based model for vacuum and in-medium hadronization. The effects of induced energy loss and nuclear absorption have been included. The main objective is the determination of the relative contribution of these mechanisms to the multiplicity ratio observable, measured in semi-inclusive deep-inelastic scattering off deuterium and nuclear targets. This is directly related to the determination of the production length, $Lp$, necessary for a quark to turn into a prehadron. We compare our results with HERMES data for multiplicity ratio and $p_t$-broadening, and show that the description of the whole data set, keeping the model parameters fixed, puts strong constrains on $Lp$. Contrary to induced-energy-loss based models, we find an important contribution from nuclear absorption at HERMES energies. Finally, we discuss some consequences of our study for the LHC physics, and we present the model predictions for the future EIC experiment.
\end{abstract}

\newpage 

\tableofcontents

\section{Introduction}
Hadronization is a complex process which can be partially described by perturbative QCD (pQCD). In the presence of a medium, this process is modified, and studying in details the modification of the hadron spectra allows the extraction of the medium properties. At typical example is the jet quenching at the LHC in Pb-Pb collisions, which can be used to estimate the size of the quark-gluon plasma (QGP). 

\begin{figure}[h]
\begin{center}
 \includegraphics[width=12pc]{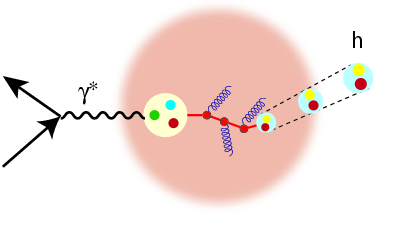}
\end{center}
\caption{\label{nuclabs} Propagation of a quark in the nucleus. After a length $Lp$, it turns into a color-singlet prehadron. During the quark propagation, gluons are emitted due to vacuum and induced energy loss.}
\end{figure}

In the opposite, a medium with known properties can be used for the study of the spacetime development of hadronization. One crucial quantity is the production length, $Lp$, corresponding to the length necessary for the quark to turn into a colorless prehadron. For the extraction of this quantity, the best observable is the hadron production in semi-inclusive deep-inelastic scattering (SIDIS) off nuclei. The idea is that, in the presence of the nuclear medium, two new phenomena appear: induced energy loss and nuclear absorption. The former is related to elastic collisions of the propagating quark, the latter to the inelastic collision of the prehadron with a nucleon, see figure \ref{nuclabs}.
These effects depend on $Lp$ and modify the hadron spectra, but not in the same way. Then, studying in details the modification of the hadron spectra allows to estimate the contributions of induced energy loss and nuclear absorption, and consequently, the extraction of $Lp$. 

In the next section, we will argue that $Lp \propto (1-z)\nu$, with $z$ the photon energy fraction carried by the hadron. There are consequently two asymptotics, leading to special cases. When $\nu\rightarrow \infty$ and $z<1$, $Lp\rightarrow \infty$. In this case, the prehadron is formed outside the nucleus and there is no nuclear absorption. This is one of the reasons why several SIDIS experiments are done at medium/low energies. The second special case is $z=1$, leaving no room for vacuum or induced energy loss, see section \ref{secielmod} for a more detailed discussion.\\

Two important observables are the $p_t$-broadening (see section \ref{ielbr}), generally attributed to the elastic collisions of the propagating quark with the medium, and the multiplicity ratio:
\begin{equation}
 R_A^h(z,Q^2,\nu)= \frac{\frac{1}{N_A^e}\frac{dN_A^h}{d\nu dzdQ^2}}{\frac{1}{N_D^e}\frac{dN_D^h}{d\nu dzdQ^2}}\simeq \frac{D_A^h(z,Q^2,\nu)}{ D^h(z,Q^2,\nu)}.\label{mult}
\end{equation}
This is the ratio of the number of hadron $h$ observed for nucleus and deuterium targets. These numbers are normalized by the respective number of electrons. The functions $D$ and $D^A$ are the in-vacuum and the in-medium fragmentation functions, respectively. At large $z$, the multiplicity ratio quantifies the hadron suppression due to the presence of the nuclear medium. These two observables have been measured in particular by the EMC \cite{EMC}, HERMES \cite{HER,HER2,HER4}, and CLAS\footnote{CLAS data are not published yet.} experiments.\\

Among the published models\footnote{Published models focus on the multiplicity ratio. It seems that we are among the first to publish a model giving a reasonnably good description of HERMES data for $p_t$-broadening.}, we can distinguish two categories. The first one is formed by induced energy loss based models,  e.g \cite{Wan,Arl}, which do not take into account the nuclear absorption. The main conclusion of these papers is that a reasonable description of SIDIS data can be achieved with induced energy loss alone, implying a negligible role of nuclear absorption. The second category contains models whose prime concern is the nuclear absorption \cite{AcMuPi,Kop,AcGrMu,PiGr,GaMo}. Note that some of these models also include induced energy loss. Their conclusions is that nuclear absorption plays an important role at moderate/low energies.\\

Energy loss and nuclear absorption are complementary, but no consensus has been reached on the respective quantitative contributions. However, answering this is necessary in order to get a numerical estimate of the production length, $Lp$, and a clearer picture of in-medium hadronization.\\

The main goals of this paper are:
\begin{enumerate}
 \item The creation of a model, including both induced energy loss and nuclear absorption, able to reproduce HERMES data for the multiplicity ratio and the $p_t$-broadening (for the leading quark, $z > 0.5$).
 \item The study of the relative contributions for these two mechanisms, as well as their kinematical dependence.
\end{enumerate}
At terms, the goal would be to make the code public. The second objective is of direct interest for the LHC physics, for instance for the $J/\psi$ production in p-Pb collision. We will discuss more in details the consequences of our study for the LHC physics in section \ref{lhcphy}.\\

The paper is structured as follows. In section \ref{model}, we present our pQCD based model for vacuum and in-medium hadronization. We give an explicit expression for the production length, $Lp$, and we discuss in details the implementation of vacuum energy loss and nuclear absorption. Section \ref{secinduced} is devoted to induced energy loss and $p_t$-broadening, which are closely related. We compare our calculations with HERMES data for $p_t$-broadening \cite{HER4} and we give more details on induced energy loss based models. In sections \ref{hermult1} and \ref{2dmultrat}, we compare our calculations with HERMES data for multiplicity ratio. We will see that the whole set of HERMES data put strong constraint on $Lp$, showing an important contribution of nuclear absorption. In our knowledge, this is the first time a theoretical paper able to describe both multiplicity ratios and $p_t$-broadening is published. In section \ref{lhcphy}, we discuss some possible implications of our work for the LHC. Finally, we show our predictions for the future EIC experiment in section \ref{eic}.

\section{Our model}\label{model}
After propagating through the nucleus on a length $Lp$, called the production length, the kicked quark turns on to a color singlet dipole. If this pre-hadronization process happens
inside the nucleus, the dipole can be ``absorbed'' by the medium, due to an inelastic collision with a nucleon. This effect leads to a reduction (enhancement) of the hadron spectra
at large (small) $z$. The fragmentation variable, $z$, is defined by:
\begin{equation}
 z=\frac{E_h}{\nu},
\end{equation}
with $E_h$ and $\nu$ the hadron and photon energies in the laboratory frame. In this section, we present our model with its assumptions, and we derive the expressions for the production length, the vacuum and the in-medium fragmentation function. We will first focus on the implementation of the nuclear absorption, leaving the discussion of induced energy loss for section \ref{secinduced}.

\subsection{pQCD based hadronization}
\begin{figure}[h]
\begin{center}
 \includegraphics[width=35pc]{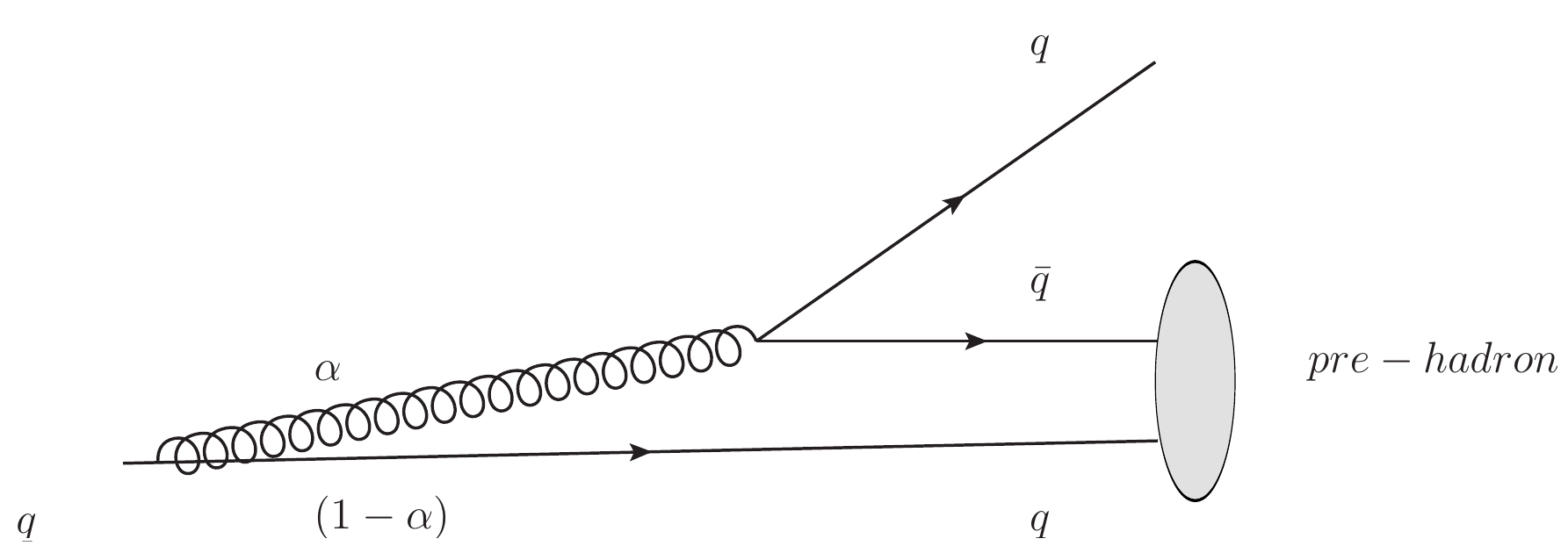}
\end{center}
\caption{\label{Be}Berger mechanism of leading pion production in Born approximation. $\alpha$ is the quark energy fraction taken by the gluon.}
\end{figure}

\begin{figure}
\begin{center}
 \includegraphics[width=35pc]{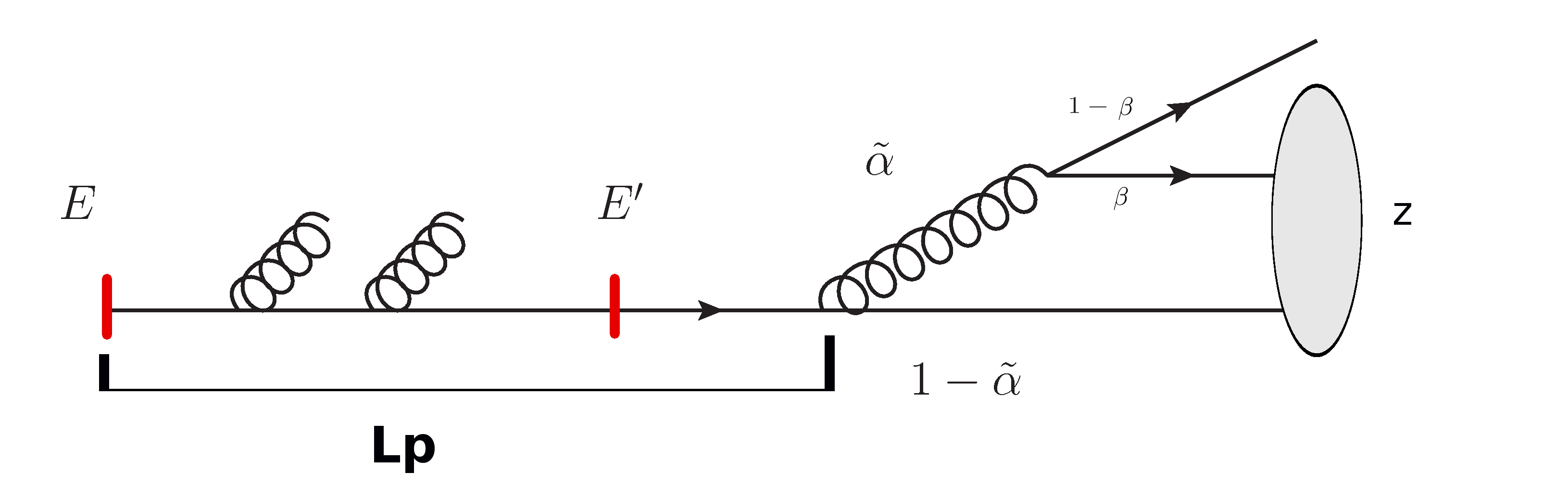}
\end{center}
\caption{\label{hober}Berger mechanism with energy loss $\Delta E = E-E'$. $\tilde{\alpha}$ is the quark energy fraction taken by gluon after energy loss $\Delta E$.}
\end{figure}

The hadronization of the leading hadron is based on the Berger model \cite{Ber}, modified by higher order considerations \cite{KoPi}. In the Born approximation, the leading quark emits a
 gluon which splits into a $q\bar{q}$ pair. The leading hadron is formed by the binding of the anti-quark the leading quark. This mechanism is illustrated in figure \ref{Be}. 
The $z$-axis is defined by the photon direction. To a good approximation, it corresponds also to the direction of the leading quark. In the limit $(1-z)\ll 1$ and $k_t\ll Q^2$, 
with $Q^2$ the usual DIS variable and $k_t$ the gluon transverse momentum relative to the photon, the pion fragmentation function reads:
 \begin{equation}
  \frac{\partial D^{Born}}{\partial k_t^2}(z,k_t^2) \propto \frac{(1-z)^2}{k_t^4}.
 \end{equation}
 In practice, we will consider $z>0.5$ as large, even if the model is expected to be completely reliable
 for larger values, $z\gtrsim 0.7$. In this equation and in the following, we use proportional rather than equal because the normalization $N$ of the fragmentation function $D$ is not written. The factor $N$ will also appear in the in-medium fragmentation function $D_A$, and will cancel in the ratio eq.~(\ref{mult}). Taking into account energy loss (see figure \ref{hober}), the pion fragmentation function at 
large z is given by \cite{KoPi}:
 \begin{equation}
   \frac{\partial D}{\partial k_t^2}(z,k_t^2) \propto \frac{(1-\tilde{z})^2}{k_t^4}.
 \end{equation}
The energy loss results in a shift of the fragmentation variable $z$ :
 \begin{equation}
  \tilde{z}=\frac{z}{1-\Delta E/E}.
 \end{equation}
Here, $E$ is the quark energy before energy loss, and $\Delta E=E-E'$ is the total energy loss. In the target frame\footnote{Which is also the laboratory frame, since we will only
consider fix-target experiments.}, the photon energy being much larger than the quark energy inside the target, we use $E=\nu$.
An important variable, in our model and for the phenomenology, is the production length $Lp$ of the color singlet dipole, which we choose to identify with the production 
length of the final gluon, splitting into the $q\bar{q}$ pair:
\begin{equation}
 Lp=\frac{E'}{M_{virt}^2} = \frac{4(1-\tilde{\alpha})\tilde{\alpha} E'}{k_t^2}
\end{equation}
The relation between $\tilde{\alpha}$ and $\tilde{z}$ is given by:
\begin{equation}
 \tilde{z}=1-\tilde{\alpha}(1-\beta), \label{ztat}
\end{equation}
with $\beta$ the energy fraction for the splitting of the gluon into the $q\bar{q}$ pair, see figure \ref{hober}. In order to keep the model simple, we will only retain the dominant kinematical 
configuration. At $\tilde{z}=1$, it is given by $\tilde{\alpha}=0$, because of the splitting function:
\begin{equation}
 P_{gq}(\tilde{\alpha})=C_F\frac{1+(1-\tilde{\alpha})^2}{\tilde{\alpha}}
\end{equation}
for the gluon emission off a quark. In the case $\tilde{z}=0.5$, equation (\ref{ztat}) imposes the lower bound $\tilde{\alpha}>0.5$. The probability for a given configuration 
is proportional to $P_{gq}(\tilde{\alpha})P_{qg}(\beta)$, where the second splitting function, for $g \rightarrow q\bar{q}$, reads:
\begin{equation}
 P_{qg}(\beta)=T_R(\beta^2+(1-\beta)^2),
\end{equation}
with $T_R=1/2$. The dominant contribution is obtained for $\tilde{\alpha}=0.5$ and $\beta=0$. Looking at the relations:
\begin{eqnarray}
\tilde{z}&=&1 \; ; \; \tilde{\alpha}=0\\
\tilde{z}&=&0.5 \; ; \; \tilde{\alpha}=0.5,
\end{eqnarray}
we deduce that:
\begin{equation}
 \tilde{z} = 1-\tilde{\alpha}. \label{zofa}
\end{equation}
While it is exact for $z\rightarrow 1$, this is a rough approximation at $z=0.5$. Inserting (\ref{zofa}) in the expression for $Lp$ gives:
\begin{equation}
 Lp = \frac{4(1-\tilde{z})\tilde{z} E'}{k_t^2}=\frac{4(1-\tilde{z})z E}{k_t^2}, \label{Lpz}
\end{equation}
where in the last equation, we used the definition of $E'$ and $\tilde{z}$ as a function of $E$ and $\Delta E$. Note that the energy loss gives a shorter production length ($\tilde{z}>z$). At $z=1$, $Lp=0$, in agreement with other estimations of the production length based on the Lund model \cite{BiGy,AcMuPi}. This behavior is expected from energy-conservation considerations. The fact that, on average, a propagating color charge looses energy, directly implies that the production length should be zero at $z=1$. In terms of $Lp$, the fragmentation function is:
\begin{equation}
 \frac{\partial D}{\partial Lp}(z,E,Q^2,Lp) \propto \frac{1- \tilde{z}}{4zE}. \label{dfrag}
\end{equation}
It depends on $Lp$ and $Q^2$ only through $\tilde{z}$. This fragmentation function could be supplemented by a Sudakov factor, similarly to what has been done in \cite{KoPi}. However, the low $Q^2$ reached at HERMES, and the partial cancellation of the Sudakov factor in the multiplicity ratio make the implementation of this function unnecessary. We checked that its implementation does not affect the results presented in this paper.

\subsection{Vacuum energy loss}\label{ven}
The perturbative contribution is given by:
\begin{equation}
 \Delta E_{pert}(L,z,Q^2) = E\int_{\lambda^2}^{Q^2}dq_t^2 \int_0^1 d\beta \beta \frac{dn_g}{dq_t^2d\beta}\Theta(L-l_c^g)\Theta(1-z-\beta),
\end{equation}
with $\lambda=0.7$ GeV a cut-off (see \cite{KoPi}), $q_t$ the gluon transverse momentum, and $\beta$ the energy fraction taken by the radiated gluon. The last step function maintains energy conservation; none of the emitted gluons can have energy bigger than $(1-z)E$.
The first step function takes into account gluons radiation time:
\begin{equation}
 l_c^g=\frac{2\beta E}{q_t^2}.
\end{equation}
A gluon can be emitted only if the quark has traveled a distance $L$ larger than $l_c^g$. The gluon number distribution reads:
\begin{equation}
 \frac{dn_g}{dq_t^2d\beta}(\beta,q_t)=\frac{2\alpha_s(q_t^2)}{3\pi}\frac{1+(1-\beta)^2}{\beta q_t^2}.
\end{equation}
After changing the variable $q_t$ for $l_c$ we have:
\begin{equation}
  \frac{dn_g}{dl_c^g d\beta}(\beta,l_c^g)=\frac{2\alpha_s(\mu^2)}{3\pi}\frac{1+(1-\beta)^2}{\beta l_c^g}, \label{dngl}
\end{equation}
with the scale in $\alpha_s$ given by:
\begin{equation}
 \mu^2=\frac{2E\beta(1-\beta)}{l_c^g}
\end{equation}
For $\alpha_s$, we use the 1-loop result, with a saturated value:
\begin{equation}
 \alpha_s(Q^2)=min\left[ \frac{12\pi}{(33-6)\ln\left(\frac{Q^2}{\lambda_{QCD}^2}\right)},0.4 \right].
\end{equation}
A saturated value for the strong coupling constant is for instance discussed in Ref. \cite{Dok}. Using Eq. (\ref{dngl}), the perturbative energy loss can be written:
\begin{equation}
 \Delta E_{pert}(L,z,Q^2) = E\int_{\lambda/E}^{1-z} d\beta \int_{l_{min}}^{l_{max}}dl  \beta \frac{dn_g}{dld\beta}, \label{enlossl}
\end{equation}
with the integration limits:
\begin{equation}
 l_{min}=\frac{2E\beta}{Q^2} \; , \;  l_{max}=min[\frac{2E\beta}{\lambda^2},L].
\end{equation}
For $L>2E\beta/\lambda^2$, none of the previous bounds depend on $L$, and the amount of perturbative energy loss stays constant. In other words, the process of perturbative energy loss stops after a distance
\begin{equation}
L_{max}=\frac{2E(1-z)}{\lambda^2},\label{Lmaxpp}
\end{equation}
where we used the fact that $\beta<1-z$.
For a more efficient numerical calculation, we apply the transformation $l\rightarrow l/2E\beta$, giving:
\begin{equation}
 l_{min}=\frac{1}{Q^2} \; , \;  l_{max}=min[\frac{1}{\lambda^2},\frac{L}{2E\beta}].
\end{equation}

Non-perturbative energy loss, related to color flux tubes, is based on the second model in Ref. \cite{KoPi}. The typical potential energy due to color flux tubes, or color strings, rises linearly with the distance. Conservation of the total energy then implies a linear decrease of the kinetic energy
\begin{equation}
\left. \frac{dE_{np}}{dL} \right|_{\textrm{string}}=-\kappa,
\end{equation}
where $\kappa$, the string tension, is taken to 1 GeV/fm. In \cite{KoPi}, a more realistic model is used, where the behavior at small distance has been modified:
\begin{equation}
\left. \frac{dE_{np}(L<E/2\lambda^2)}{dL} \right|_{\textrm{string}}=-\frac{2\lambda^2}{E}L\kappa,
\end{equation}
 leading to:
\begin{equation}
 \Delta E_{np}(L,z) =\kappa \frac{\lambda^2 L^2}{E}\Theta(L_{max}-L) + \left[\kappa \frac{\lambda^2 L_{max}^2}{E}+\kappa (L-L_{max})\right]\Theta(L-L_{max}),
\end{equation}
with $L_{max}$ defined in Eq. (\ref{Lmaxpp}).
The vacuum energy loss is given by the sum of the perturbative and non-perturbative contributions. We stop the process when $\Delta E$ reaches the values of $(1-z)E$. The induced energy loss will be discussed later, in section \ref{ielbr}.

\subsection{Fragmentation function}
\begin{figure}
\centering
\includegraphics[width=10cm]{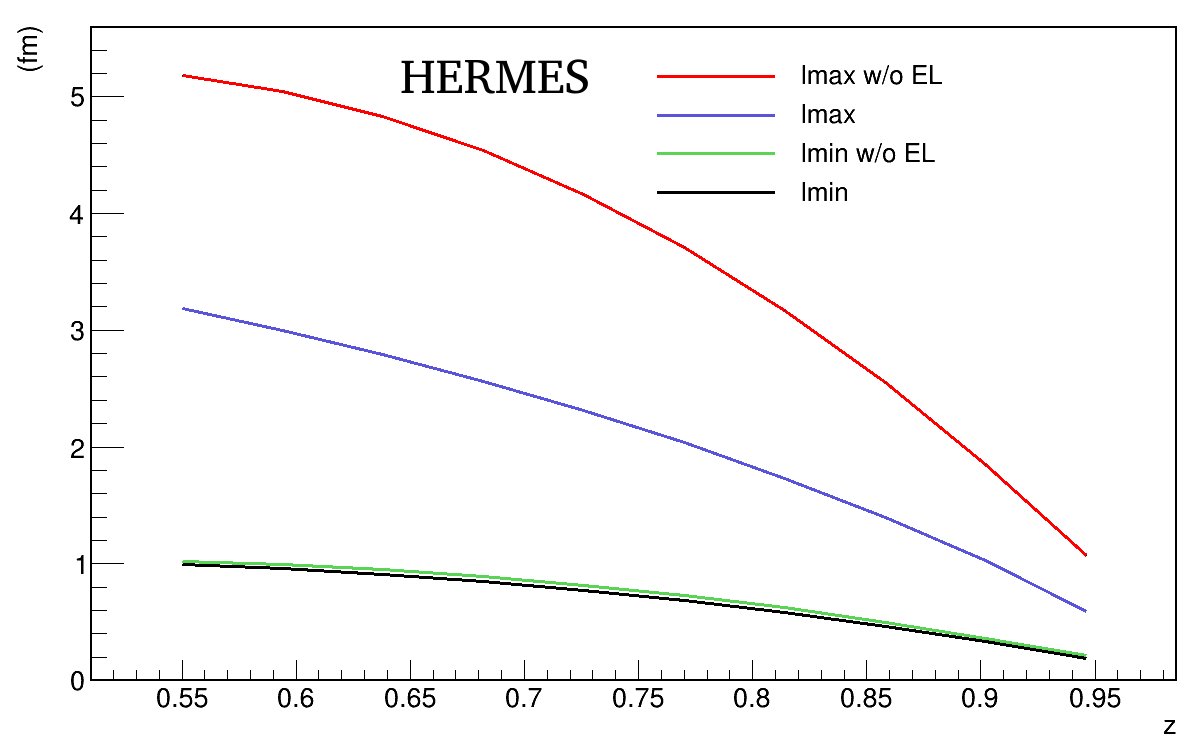}
\caption{$Lp_{max}$ and $Lp_{min}$, in fm, with and without vacuum energy loss. The effect of energy
loss is larger for $Lp_{max}$. We used $E=13$ GeV and $Q^2=2.5$ GeV$^2$.\label{lp}}
\end{figure}

In order to compare with data, one has to integrate the differential fragmentation function (\ref{dfrag}) over the production length $Lp$ :
\begin{equation}
D(z,Q^2,E)\propto \int_{Lp_{min}}^{Lp_{max}} dLp \frac{\partial D}{\partial Lp}(z,Q^2,E,Lp), \label{fragvac}
\end{equation}
with $Lp_{min}$ and $Lp_{max}$ given by the equations :
\begin{equation}
Lp_{min}=\frac{4Ez(1-\tilde{z}(Lp_{min}))}{Q^2} \; ; \; Lp_{max}=\frac{4Ez(1-\tilde{z}(Lp_{max}))}{\lambda^2}.
\end{equation}
Since $\tilde{z}<z$, we see that the production length is shorted by energy loss. These equations are solved numerically and solutions are plotted as a function of $z$ in 
figure \ref{lp}. In the same figure, we show the solution in the Born approximation (no energy loss).
As expected, everything goes to zero when $z\rightarrow 1$.\\

\begin{figure}
\centering
\includegraphics[width=10cm]{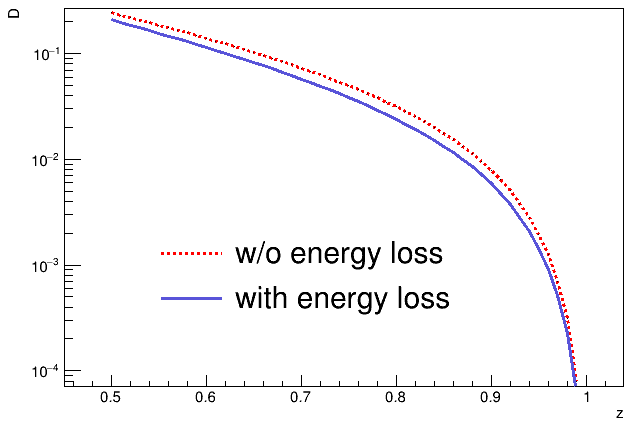}
\caption{Fragmentation function with and without energy loss. We used $E=13$ GeV and $Q^2=2.5$ GeV$^2$. The absolute normalization is not computed, this factor will cancel in the ratio.}\label{dpi}
\end{figure}

Figure \ref{dpi} displays the fragmentation function (\ref{fragvac}) obtained with and without energy loss. The absolute normalization has not been computed. We
observe that with energy loss, the slope of the fragmentation function is softer.
Quantitatively, the effect does not seem to be important for the vacuum fragmentation function.

\subsection{In-medium hadronization}
The nuclear fragmentation function is simply the convolution of the vacuum fragmentation function, equation (\ref{dfrag}), with a suppression factor :
\begin{equation}
 D^A(z,Q^2,E) \propto \frac{1}{A}\int d^2 b \int dz_l \, \rho(b,z_l) \int_{Lp_{min}}^{Lp_{max}}dLp \frac{\partial D}{\partial Lp}(z,Q^2,E,z_l,z_l+Lp)Tr(z,Q^2,E,b,z_l+Lp,\infty).\label{nuclfra}
\end{equation}
Here $b$ is the two dimensional impact parameter and $\rho$ the nuclear density, taken from \cite{density}. $z_l$ is the longitudinal coordinate of the DIS process and $Tr$ the suppression factor due to dipole 
absorption by the nuclear medium. Our calculations include the induced energy loss, Eq.(\ref{ieltotal}), due to the quark propagation from $z_l$ to $z_l+Lp$. It explains the dependence of the fragmentation function $\partial D/\partial Lp$ on $z_l$ in Eq. (\ref{nuclfra}) (this variable is absent in Eq. (\ref{dfrag}), for the in-vacuum case). Then, the dipole is formed at $z_l+Lp$ and can be absorbed during its propagation, with a probability which depends of course on the nuclear density. 
In the approximation where the dipole size is frozen during its travel throughout the nucleus, the survival probability, which we also call
nuclear transparency, is given by \cite{KoZa} :
\begin{equation}
 Tr(z,Q^2,E,b,z_1,z_2)=\left| \frac{\left< \psi_h(r_t)|\exp[-(1/2)\sigma_{\overline{q} q}(r_t)T_A(b,z_1,z_2)]|\psi_{\overline{q}q}(r_t)  \right>}{\left< \psi_h(r_t)|\psi_{\overline{q}q}(r_t) \right>} \right|^2.
\end{equation}
The explicit $z$, $Q^2$ and $E$ dependence is not shown in order to keep the equation readable. It enters through the dipole cross section, $\sigma_{\overline{q} q}$, and dipole wave 
function $\psi_{\overline{q}q}$. $z_1$ and $z_2$ are longitudinal coordinates and $r_t$ is a two dimensional vector for the dipole size. $T_A$, the thickness function, is given by :
\begin{equation}
 T_A(b,z_1,z_2)=\int_{z_1}^{z_2} dl \, \rho(b,l).
\end{equation}
At low energy, the dipole size can fluctuate and the $q\overline{q}$ propagation throughout the medium is achieved with the light cone Green function 
$G(z_2,\overrightarrow{r_2},z_1,\overrightarrow{r_1})$. The variables $z_1$, $z_2$ correspond to initial and final times, respectively, whereas $\overrightarrow{r_1}$, $\overrightarrow{r_2}$ represent the 
initial and final dipole sizes. This Green function obeys to the two dimensional light cone Schrödinger equation, described in \cite{KoScTa}:
\begin{equation}
 i\frac{d}{dz_2}G(z_2,\overrightarrow{r_2},z_1,\overrightarrow{r_1})=\left[ \frac{\epsilon^2-\Delta_{r_2}}{2p\beta(1-\beta)} +V_{q\overline{q}}(z_2,\overrightarrow{r_2},\beta)\right]G(z_2,\overrightarrow{r_2},z_1,\overrightarrow{r_1}) \label{schro},
\end{equation}
with $\beta$ the light cone fraction of the quark inside the pion wave function. Using the Green function we have for the transparency factor:
\begin{equation}
 Tr(z,Q^2,E,b,z_1,z_2)=\left| \frac{\int d^2r_1 d^2r_2 \psi_h^*(r_2)G(z_2,r_2,z_1,r_1)\psi_{q\overline{q}}(r_1)}{\int d^2r \psi_h^*(r)\psi_{q\overline{q}}(r)} \right|^2 \label{coltr}.
\end{equation}
For the hadronic wave function, we use a parametrization in the form of the asymptotic light-cone meson wave function \cite{KoNeSc} :
\begin{equation}
 \psi_h(r)=f(\beta)\exp(-a^2(\beta)r^2/2), \label{psih}
\end{equation}
with:
\begin{equation}
 a^2(\beta) = \frac{4(\beta(1-\beta)+a_0)}{\left<r_{\pi}^2 \right>}. \label{a2bbeta}
\end{equation}
The authors of Ref.  \cite{Kop} found that for the value $a_0=1/12$, the previous wave function reproduces correctly the pion mean radius squared.
In our model, the light cone fraction, $\beta$, of the quark inside the pion wave function is related to $z$ by:
\begin{equation}
 \beta=\frac{2\widetilde{z}-1}{\widetilde{z}}. \label{beta}
\end{equation}
The wave function (\ref{psih}) is solution of the Schrödinger equation (\ref{schro}) if the real part of the potential is given by:
\begin{equation}
 \Re \; V_{q\overline{q}}(z_2,\overrightarrow{r},\beta)=\frac{a^4(\beta)r^2}{2p\beta(1-\beta)}. \label{repot}
\end{equation}
The imaginary part of the potential, responsible for the dipole absorption, reads:
\begin{equation}
 \Im \; V_{q\overline{q}}(z_2,\overrightarrow{r},s)= -\frac{\sigma_{q\overline{q}}(s,r)}{2}\rho_A(z_2), \label{impot}
\end{equation}
with $\sqrt{s}$ the pre-hadron nucleon center of mass energy, and $\sigma_{q\overline{q}}(r)$ the dipole-nucleon cross section:
\begin{eqnarray}
 \sigma_{q\overline{q}}(s,r) &=& C(s)r^2,\\
 s&=& 2zM_pE+M_p^2+m_{h}^2,
\end{eqnarray}
where for pratical calculations, the pre-hadron mass has been identified with the one of the detected hadron. The expression for the energy dependent factor $C$ has been taken from \cite{KoPoSc}:
\begin{eqnarray}
C(s) &=& \frac{1}{4}\sigma_{tot}^{\pi p}(s)\left[Q^2_{qN}(s)+\frac{3}{2\left<r^2_{ch}\right>_{\pi}}\right],\label{cx}\\
Q_{qN}(s) &=& 0.19 \,\mbox{GeV} \times \left(\frac{\sqrt{s}}{1 \,\mbox{GeV} }\right)^{0.14}.
\end{eqnarray}
Here we use $\sigma_{tot}^{\pi p}(s)=23.6\left( s/s_0 \right)^{0.079}+1.432\left( s/s_0 \right)^{-0.45}$ mb \cite{Bar} with $s_0=1000$ GeV$^2$ for $\sqrt{s}>2.5$ GeV$^2$ and a
 table from Igor Strakovsky otherwise.
Moreover, we use the relation $\left<r_{\pi}^2 \right> = \frac{8}{3}\left<r_{\pi}^2 \right>_{em}$, with $\left<r_{\pi}^2 \right>_{em}=0.44$ fm$^2$ being the pion mean charge 
radius squared \cite{Ame}. With this potential, Eqs. (\ref{repot}) and (\ref{impot}), one obtains a Schrödinger equation for an harmonic oscillator with the frequency:
\begin{equation}
 \xi=\frac{\sqrt{a^4(\beta)-i\mu C(s)\rho(\overrightarrow{r})}}{\mu} \; ; \; \mu=E_h\beta(1-\beta). \label{eqxi}
\end{equation}
The solution of this equation is \cite{KoScTa}:
\begin{equation}
G(z_2,\overrightarrow{r_2},z_1,\overrightarrow{r_1})=\frac{A}{2\pi i \sin(\xi\Delta z)}
 \exp \left(-\frac{A}{2}\left[(r_1^2+r_2^2)\frac{\cos(\xi\Delta z)}{i \sin(\xi\Delta z)}-\frac{2\overrightarrow{r_1}.\overrightarrow{r_2}}{i \sin(\xi\Delta z)} \right] \right), \label{grefun}
\end{equation}
where $\Delta z=z_2-z_1$ and $A=\mu \xi$.\\

The last ingredient for the nuclear transparency factor, Eq. (\ref{coltr}), is the dipole wave function. Because we want a continuous transition between the dipole and pion
wave functions, we will use:
\begin{equation}
 \psi_{\overline{q}q}(z,Q^2,E,Lp,r)=f(\beta)\exp(-b^2(\beta)r^2/2), \label{psiq}
\end{equation}
with:
\begin{equation}
 b^2(\beta)=\frac{4(\beta(1-\beta)+a_0)}{\left< r^2_{q\overline{q}} \right>}.
\end{equation}
Then, for $ \left< r^2_{q\overline{q}} \right> = \left< r^2_{\pi} \right>$, the dipole and pion wave functions are equal.
The dependence on $z,Q^2,E,Lp$ enters through the dipole mean radius and is described in Sec. \ref{secevol}.\\

Using Eqs. (\ref{psih}), (\ref{grefun}) and (\ref{psiq}), we find that for a constant nuclear density:
\begin{equation}
 Tr(z,Q^2,E,b,z_1,z_2)= \left| \frac{a^2(\beta)+ b^2(\beta)}{2\frac{B}{A}i\sin(\xi \Delta z)a^2(\beta)+2B\cos(\xi \Delta z)-\frac{A}{i\sin(\xi \Delta z)}} \right|^2,
\end{equation}
with :
\begin{equation}
 B=\frac{1}{2}\left(A\frac{\cos(\xi \Delta z)}{i\sin(\xi \Delta z)})+ b^2(\beta)\right).
\end{equation}
In the case $C(s)=0$ (no nuclear absorption), Eq.(\ref{eqxi}) gives $\mu \xi=A=a^2(\beta)$, and we can check explicitly that $Tr=1$. For a non-constant nuclear density, we have to
discretize the time (or equivalently the space in the longitudinal direction), and the nuclear transparency factor is given by :
\begin{equation}
 Tr(z,Q^2,E,b,z_1,z_2)=\frac{\pi}{ b^2(\beta)}( b^2(\beta)+a^2(\beta))^2\left| \frac{F_n}{A_n+a^2(\beta)}\right|^2, \label{trd}
\end{equation}
with the following recurrence relations :
\begin{align}
b_n &=\frac{\mu \xi_n}{i\sin(dz \xi_n)}\\
A_n &=b_n\left(\cos(dz \xi_n)-\frac{1}{\cos(dz \xi_n)+A_{n-1}/b_n}\right)\\
F_n &=F_{n-1}\left(\cos(dz \xi_n)-\frac{A_n}{b_n}\right)\\
F_0 &=\sqrt{\frac{ b^2(\beta)}{\pi}}\\
A_0 &= b^2(\beta),
\end{align}
where $n$ is the number of steps and $dz=\Delta z/n$. In the case $n=0$ (no propagation), $Tr=1$. Our numerical results, to be presented later, have been obtained with $N=1000$.

\subsection{Dipole size evolution at $l<Lp$}\label{secevol}
The missing ingredient for the computation of the nuclear transparency factor, is the dipole mean radius squared $\left< r^2_{q\overline{q}} \right>$, entering through the definition
of the dipole wave function, Eq. (\ref{psiq}). In our model, the dipole is \textit{really} produced at $l=Lp$, which corresponds to the moment when the medium is able to resolve the 
dipole and the accompanying quark separately. For $l>Lp$, the dipole evolution is entirely managed by the Schrödinger equation, Eq. (\ref{schro}), and $\left< r^2_{q\overline{q}} \right>(Lp)$
is an initial condition for this evolution. For $l<Lp$, there is an evolution of the dipole transverse size, but the dipole cannot be absorbed since it cannot be resolved
by the medium. In this case, in agreement with \cite{GaMo} and \cite{Dok2}, we choose the dipole cross section to rise linearly with $l$. The dipole cross section being proportional to
 its squared transverse size, we have:
\begin{equation}
 \frac{\left< r^2_{q\overline{q}} \right>(Lp,Q^2,E,z)}{\left< r^2_{\pi} \right>} = x_0(Q^2) + [1-x_0(Q^2)]\frac{Lp}{t_f(z,E)}, \label{dipevo}
\end{equation}
with $t_f$ the formation time and $Lp\leq t_f$. The normalization has been chosen such that $\left< r^2_{q\overline{q}} \right>=\left< r^2_{\pi} \right>$ when $Lp=t_f$. Indeed, at $t=t_f$, the wave function of the hadron is fully formed and the transverse size of the $q\bar{q}$ bound state should correspond to the hadron transverse size. The initial dipole size (at $l=0$) is $r_0^2= r_{\pi}^2 x_0$, with $x_0<1$. Our choice for these parameters is:
\begin{eqnarray}
x_0&=&\frac{0.05}{Q^2}\label{x0}, \\
t_f &=& a_f \frac{zE}{2\lambda^2}\label{af},
\end{eqnarray}
with $a_f=0.9$ and $\lambda^2$ the cutoff used in section \ref{ven}. The $1/Q^2$ dependence of $r_0^2$ is expected from phenomenology and is responsible for color transparency.
The $zE$ behavior of the formation length is the consequence of a Lorentz boost to energy $E_h=zE$. The numerical values of these two parameters have been chosen in order to 
improve our results for the multiplicity ratio. However, we will see that numerical calculations show little dependence on $x_0$, leaving effectively one free parameter.\\

\section{Induced energy loss}\label{secinduced}

\subsection{$p_t$-broadening and induced energy loss}\label{ielbr}
Experimentally, the $p_t$-broadening is defined as the difference between the mean transverse momentum squared measured for the proton/deuterium target and a nuclear target:
\begin{equation}
\Delta_{\textrm{exp}} \left<p_t^2\right>_h= \left<p_t^2 \right>_h^A-\left<p_t^2 \right>_h^p,
\end{equation}
with the subscript $h$ referring to the hadron species. It is generally considered that the main contribution to the $p_t$-broadening is the induced energy loss, due to the quark propagation through the nucleus. The two main formulas \cite{JoKoTa} are:
\begin{equation}
 \Delta \left<p_t^2 \right>(s,z_1,b,Lp)=2C(s)\int_{z_1}^{z_1+Lp} \rho(l,b) dl, \label{ptb}
\end{equation}
with $C(s)$ given in Eq. (\ref{cx}), and $z_1$ the longitudinal coordinate of the DIS interaction, and \cite{BDMPS}
 \begin{equation}
 \Delta E_{iel}(s,z_1,b,Lp) = \frac{3}{4}\alpha_s \int_0^{Lp} \Delta \left<p_t^2 \right>(s,z_1,b,l)dl. \label{ieltotal}
\end{equation}
The last equation gives the total amount of induced energy loss due to the propagation of the quark in the nuclear medium.\\

\begin{figure}
\centering
\includegraphics[width=12cm]{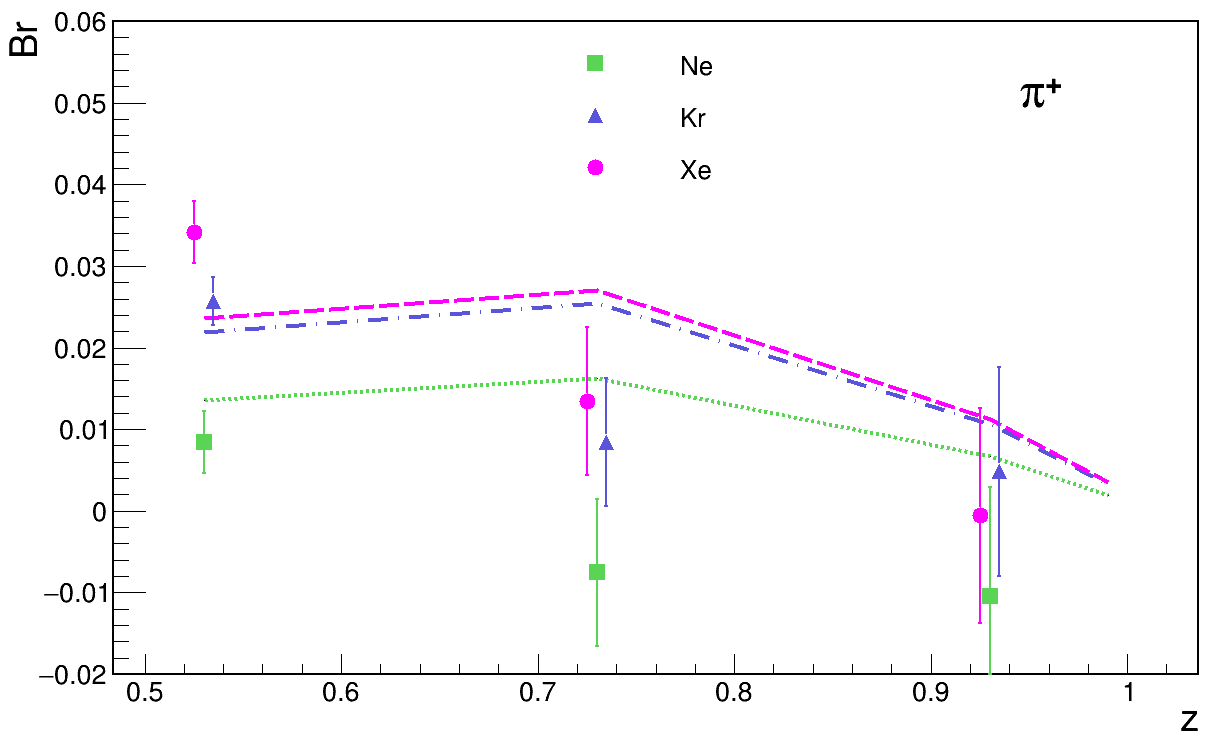}
\caption{$\pi^+$ $p_t$-broadening ("Br") for Hermes experiment \cite{HER4}}\label{brher2}
\end{figure}

The DIS interaction can occur everywhere in the nucleus, and we have to average Eq. (\ref{ptb}) with $\frac{1}{A}\int d^2b \, dz_1 \, \rho(z_1,b)$. As usual, the average on $Lp$ is obtained using the differential fragmentation function, $\frac{1}{N_L}\int dLp \frac{\partial D}{\partial Lp}$. All together, the quark $p_t$-broadening is given by:
\begin{equation}
\Delta \left<p_t^2 \right>(z,Q^2,E)=\frac{2C}{A}\int d^2b \int dz_1 \rho(z_1,b)\int_{L_{min}}^{L_{max}}dLp\frac{1}{N_L}\frac{\partial D}{\partial Lp}T_A(b,z_1,z_1+Lp),
\end{equation}
with 
\begin{equation}
T_A(b,z_1,z_1+Lp)=\int_{z_1}^{z_1+Lp} dl \rho(l,b),
\end{equation}
and the normalization
\begin{equation}
N_L(z,Q^2,E) = \int_{L_{min}}^{L_{max}}dLp \frac{\partial D}{\partial Lp}(z,Q^2,E,Lp).
\end{equation}
Finally, the quark $p_t$-broadening is related to the hadron $p_t$-broadening by:
\begin{equation}
 \Delta \left<p_t^2 \right>_h = \tilde{z}^2\Delta \left<p_t^2 \right>. \label{brhh}
\end{equation}
In Fig. \ref{brher2}, our calculations are compared to HERMES data \cite{HER4}.
We can see that in the second and third bin, the data points are below our calculations, or even below zero. We believe that it might be explained by an interesting physical effect which has not been considered in the literature, but we keep these considerations for another publication.\\

At $z=1$, the $p_t$-broadening (due to induced energy loss) goes to zero due to energy conservation. In our calculations, the slight increase between the first and second bin is due to the $z^2$ factor in Eq. (\ref{brhh}).

\subsection{Note on induced-energy-loss based models}\label{secielmod}
In the next section, we will discuss our results for the multiplicity ratio. Before this discussion is in order, it is useful to present the idea of the models based on induced energy loss, e.g. \cite{Wan,Arl}, inspired by Ref \cite{Wang2}. In all these models (as well as in ours), the fragmentation variable, $z$, is shifted to higher values due to induced energy loss. 
The fragmentation function decreasing with $z$, this mechanism results in a suppression of the multiplicity ratio Eq. (\ref{mult}). We illustrate this mechanism by the model used
in \cite{Arl}, where the nuclear fragmentation function is given by:
\begin{equation}
 zD_q^h(z,Q^2,A)=\int_0^{\nu-E_h=(1-z)\nu}d\epsilon D(\epsilon,\nu)z^*D_q^h(z^*,Q^2), \label{ielmod}
\end{equation}
with $D_q^h(z,Q^2)$ the vacuum fragmentation function of the quark $q$ into the hadron $h$, and $z^*=1/(1-\epsilon/\nu)$. The quenching weight \cite{BDMS}, $D(\epsilon,\nu)$, gives the probability distribution for a quark of energy $\nu$ to loose an energy $\epsilon$. It depends on $Lp$, the distance covered by the quark before its prehadronization into a colorless dipole:
\begin{equation}
D(\epsilon,\nu,Lp)=\sum_{n=0}^{\infty}\frac{1}{n!}\left[\prod_{i=1}^n \int d\omega_i \frac{dI(\omega_i,\nu,Lp)}{d\omega_i}\right]\delta\left(\epsilon-\sum_{i=1}^n \omega_i\right)\exp\left[-\int_0^{\infty}d\omega \frac{dI(\omega,\nu,Lp)}{d\omega}\right]. \label{quench}
\end{equation}
The gluon spectrum, $dI/d\omega$, radiated by hard quarks produced in QCD media has been computed in \cite{BDMPS}. In the first order in quark energy, $\mathcal{O}(1/\nu)$, it reads:
\begin{equation}
\frac{dI}{d\omega}(\omega,\nu,Lp)=\frac{\alpha_s C_F}{\pi\omega}\left(1-\frac{\omega}{\nu}\right)\ln\left[\cosh^2\sqrt{\frac{\omega_c}{2\omega}}-\sin^2\sqrt{\frac{\omega_c}{2\omega}}\right]\Theta(\nu-\omega)\label{iomeg},
\end{equation}
and the dependence on $Lp$ enters through the variable:
\begin{equation}
\omega_c=\frac{1}{2}\hat{q}Lp^2.
\end{equation}
For $Lp$ larger than the nuclear size, the suppression of the hadronic spectrum is entirely due to induced energy loss, the dipole being formed outside of the nuclei. In the opposite,
at very small $Lp$ ($z$ close to 1), the pre-hadron is formed (nearly) instantaneously and the nuclear absorption is maximum. In this situation, a simple physical argument: $Lp=0\Rightarrow$ no quark energy loss $\Rightarrow \epsilon=0$, shows that $D(\epsilon,\nu)\rightarrow \delta(\epsilon)$. Mathematically, it can be seen noting that $dI/d\omega$ behaves like $\delta(\omega)$, when $Lp=0$. Indeed, for $\omega_c=0$ and $\omega \neq 0$, the result of Eq.(\ref{iomeg}) is zero due to the logarithm. If $\omega_c=\omega=0$, the logarithm is not zero and the function is divergent due to the factor $1/\omega$. Changing $dI/d\omega$ by $\delta(\omega)$ in Eq. (\ref{quench}) gives:
\begin{equation}
D(\epsilon,\nu,Lp)=\sum_{n=0}^{\infty}\frac{1}{n!}\delta(\epsilon)\exp(-1),
\end{equation}
where $\sum_i \omega_i=0$ has been used. Finally, noting that $\sum_{n=0}^{\infty}\frac{1}{n!}=\exp(1)$, we get that
\begin{equation}
\lim_{Lp\to 0} D(\epsilon,\nu,Lp) = \delta(\epsilon).
\end{equation}
In this limit, IEL based models predict a multiplicity ratio equal to 1, as can be seen directly from Eqs. (\ref{ielmod}) and (\ref{mult}). In other words, the suppression of the hadronic spectrum is entirely due to nuclear absorption. More generally, any IEL based model obeying energy conservation, $\epsilon=0$ if $z=1$, should predict a multiplicity ratio equal to 1 at $z=1$.\footnote{In general, IEL based models show their result with a cut for $z>0.95$, see for instance \cite{Wan,Arl}. In this papers, it can be seen that their predictions for the multiplicity ratio starts to increase at $z\sim 0.85$.}\\

The interest of such models based on the quenching weight is their simplicity. However, they can't be used in realistic situations due to the divergences of the quenching weight when $\epsilon \rightarrow 0$, corresponding to a very small amount of energy loss. It happens for instance at small nuclear density. For this reason, one has to use a hard sphere model for the nucleus, with constant density, and forbid the DIS interaction to occur too close to the back edge.

\section{Results for the HERMES multiplicity ratio : $\pi^+$}\label{hermult1}
\begin{figure}[!h]
\centering
\includegraphics[width=12cm]{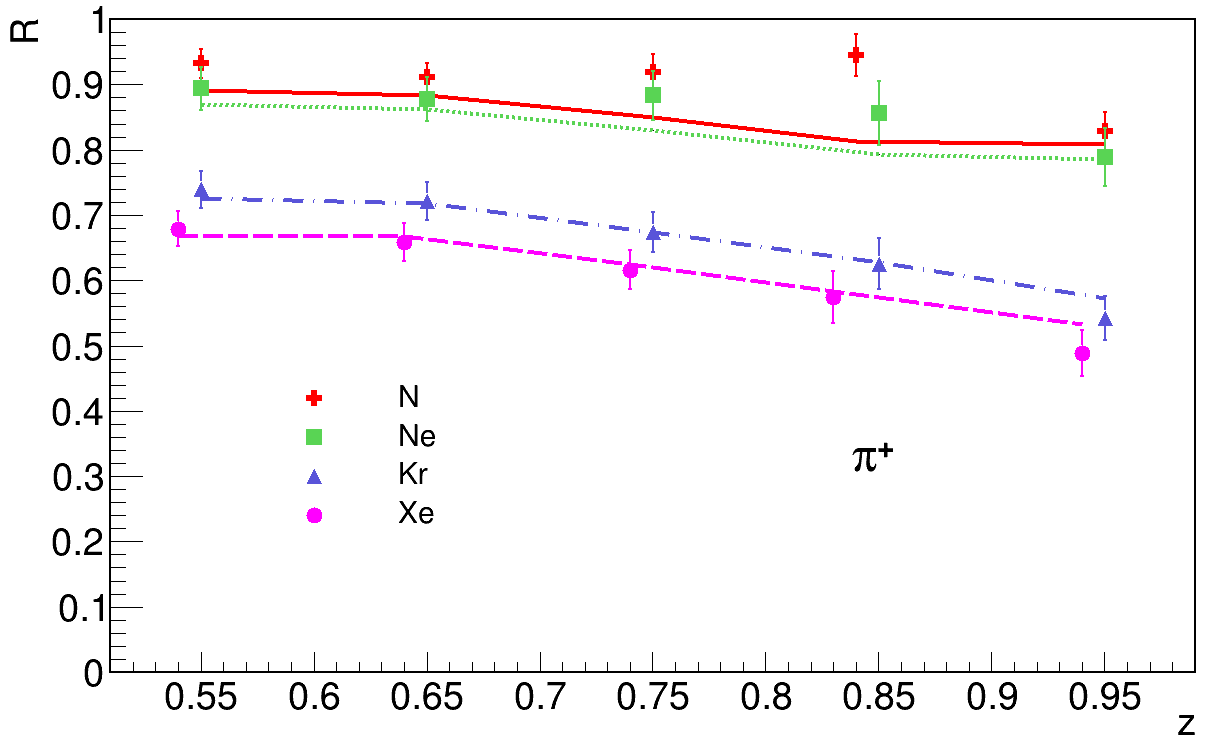}
\caption{ Multiplicity ratio for $\pi^+$, compared to HERMES data \cite{HER,HER2}. The solid line is for nitrogen, the dotted line for neon, the dashed-dotted line for krypton and the dashed line for xeon. \label{her1}}
\end{figure}

We now have all the necessary ingredients for the computation of the multiplicity ratio, which, to a good approximation, is given by Eq. (\ref{mult}).
Our nuclear fragmentation function, Eq. (\ref{nuclfra}), also include the induced energy loss via the shift of the variable $z$:
\begin{equation}
 \tilde{z} = \frac{z}{1-(\Delta E_{vac}+\Delta E_{iel})/E}. \label{ztild2}
\end{equation}
The total energy loss is not allowed to be larger than $(1-z)E$. Then, the variable $\tilde{z}$ is always smaller than 1. In Fig. \ref{her1}, we show the comparison between our calculations and HERMES data \cite{HER,HER2} for the $\pi^+$ particle.
For heavy nuclei, our calculations are in perfect agreement with experimental data. For light nuclei, the results are also quite satisfying, but it seems that we are slightly undershooting the data. We note that at $z=0.85$, the data for nitrogen goes up, which could be due to some systematics.\footnote{However, the same is observed for CLAS data on the carbon nucleus.}\\

\begin{figure}[!h]
\centering
\includegraphics[width=8.3cm]{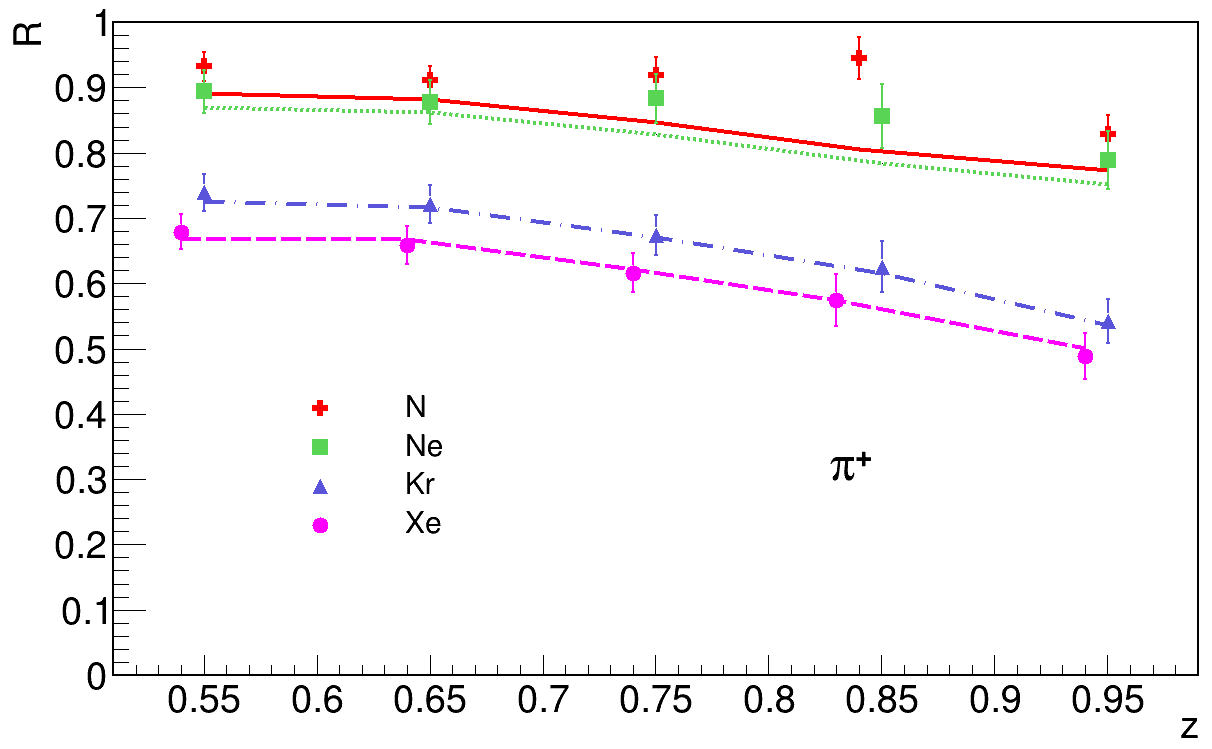}
\includegraphics[width=8.3cm]{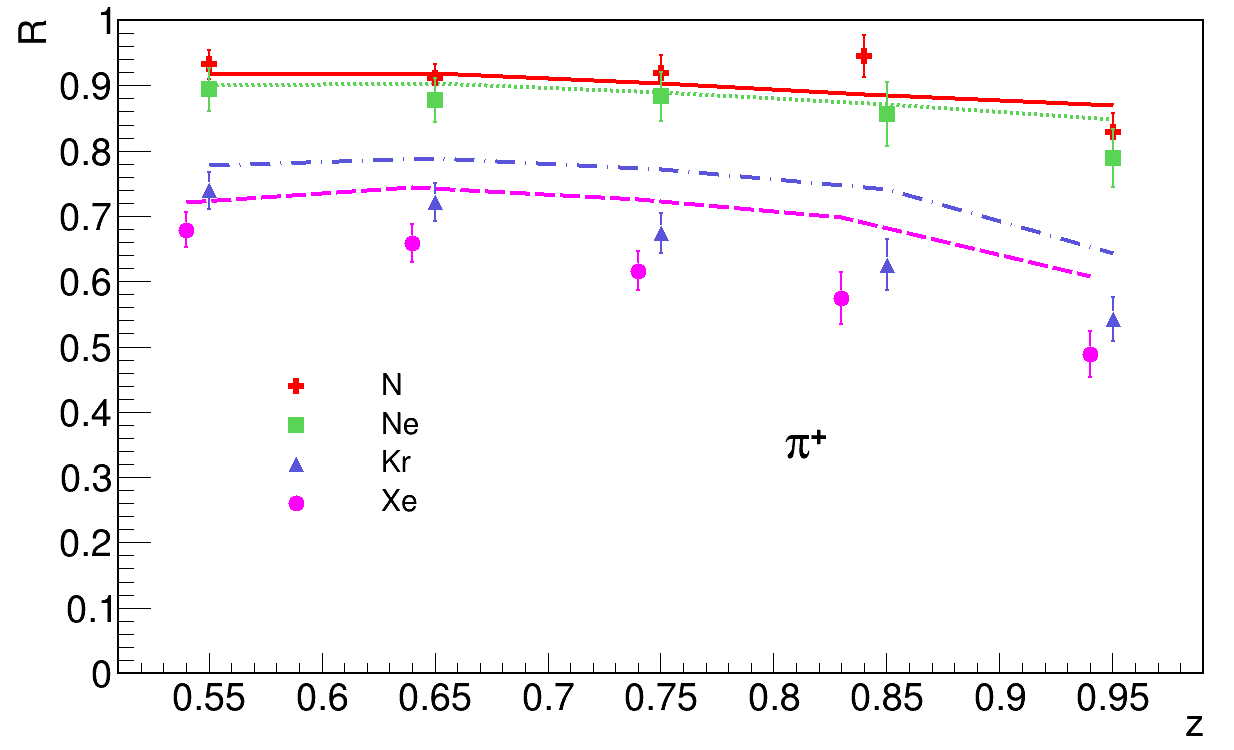}
\caption{Left: same as Fig. \ref{her1} with $x_0$ [see Eq. (\ref{x0})] multiplied by 5. Right: same as Fig. \ref{her1} with $a_f$ changed from 0.9 to 2.5.\label{herm2}}
\end{figure}

In Fig. \ref{herm2}, we show the dependence of our results on the parameters $x_0$ and $a_f$, see Eqs. (\ref{x0}) and (\ref{af}).
Multiplying $x_0$ by 5 has only a visible effect in the last bin. In the opposite, the calculations show more dependence on $a_f$. Globally, we see a smooth and reasonable dependence of our results on the model parameters.\\

\begin{figure}[!h]
\centering
\includegraphics[width=12cm]{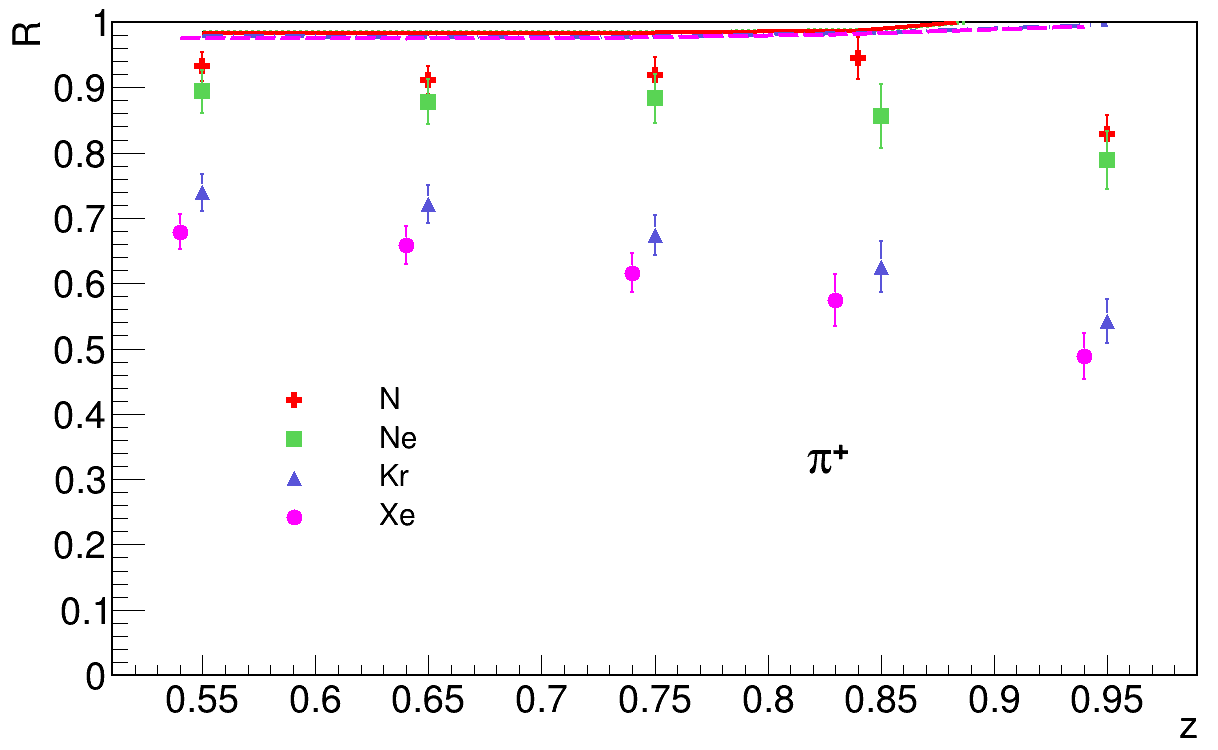}
\caption{Same as Fig. \ref{her1}, taking into account only the effect of induced energy loss (nuclear absorption turned off).\label{hereind}}
\end{figure}

In order to study the contribution of induced energy loss (IEL), we switch off the nuclear absorption, giving the result shown in Fig. \ref{hereind}.
This contribution depends of course on several variables like $z$ or the photon energy. We can see that, as expected, IEL gives no suppression at $z=1$. For heavy nuclei, the IEL contribution is approximately 7$\%$ at $z=0.55$, while for light nuclei, the value is at least 2 times larger.\\

It could looks like we are underestimated the IEL contribution (for a given $Lp$). But it is probably not the case for the following reasons. First, our implementation of IEL plus nuclear absorption gives a good description of the data. Moreover, at $z=0.55$, the nuclear absorption  depends weakly on the 2 free parameters, as shown in Fig. \ref{herm2}, leaving nearly no choice for the amount of IEL. Second, the IEL is related to the $p_t$-broadening, and our calculations give a satisfying description of HERMES data, Fig. \ref{brher2}.\\

\begin{figure}[!h]
\centering
\includegraphics[width=8.7cm]{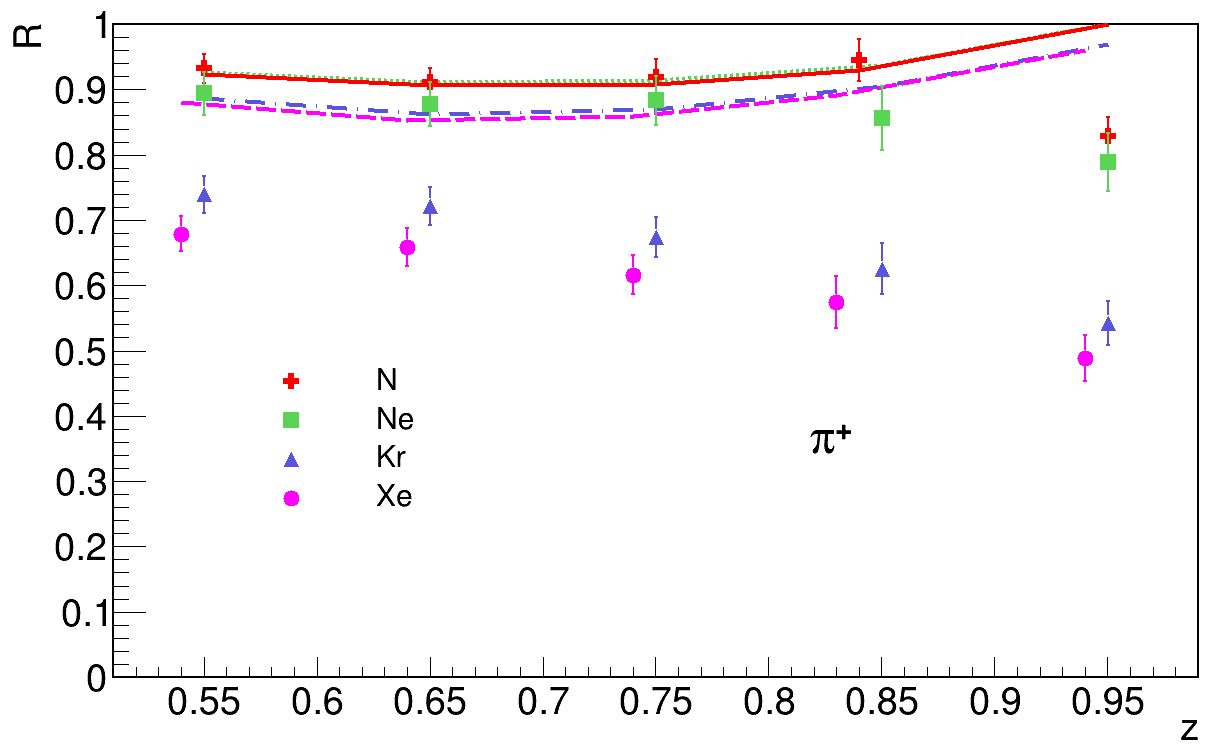}
\includegraphics[width=8.0cm,height=5.4cm]{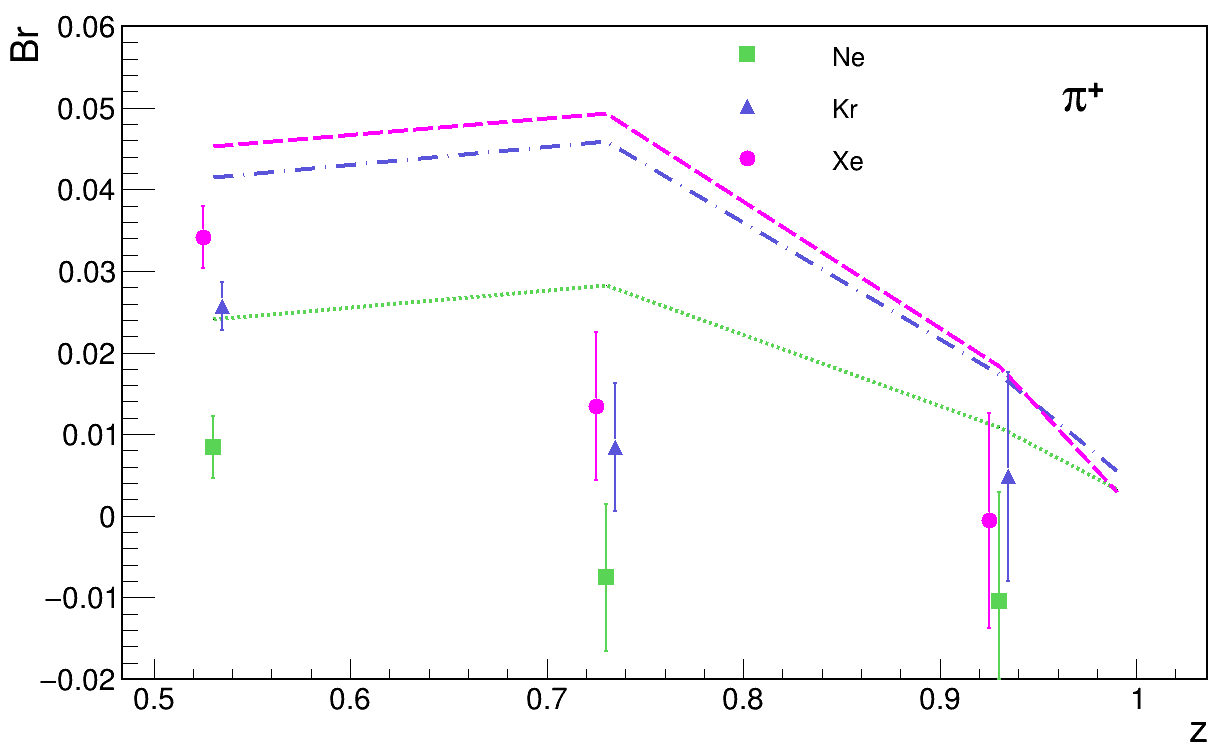}
\caption{Left: contribution of IEL to the multiplicity ratio if we multiply our $Lp$ by 2 (nuclear absorption switched off). Right: the corresponding $p_t$-broadening compared to HERMES data.\label{herm3}}
\end{figure}

One could also be skeptical on our estimation of the production length $Lp$. A larger $Lp$ will increase the IEL contribution and decrease the nuclear absorption contribution. It is not at all obvious that the change in IEL will compensate the change in nuclear absorption, but in fact, in some extent, it does. Multiplying $Lp_{min}$ and $Lp_{max}$ by a factor 2, and adjusting the parameter $a_f$ (to 2.5), we observed that our model gives still a good description of HERMES data for the multiplicity ratio. In this case, the contribution of induced energy loss is significantly increased, see Fig. \ref{herm3} (left).
However, with this larger $Lp$, the result for $p_t$-broadening overestimates HERMES data, see Fig. \ref{herm3} (right).\\

Note that in the left panel of Fig. \ref{herm3}, the IEL alone does a good description of 4 data points for nitrogen. However, this effect alone does not give the correct shape, and it is impossible to be in agreement with all data points. This situation is exactly what happens in \cite{Arl}, figure 7. Moreover, the estimation by the author of IEL is even larger than ours, after multiplying $Lp$ by 2. It implies that the corresponding theoretical $p_t$-broadening will completely overshoot HERMES data. The fact that, in the past, some studies based only on IEL claimed that: 1) it is possible to reproduce experimental data just with IEL; and 2) the contribution of nuclear absorption is small; has probably some consequences for the LHC physics (see section \ref{lhcphy} for more details). For this reason, it is important to understand the different contributions in SIDIS experiments, and we will now summarize the conclusions obtained in this section.\\

Our model gives a good description of HERMES data for multiplicity ratio\footnote{See the next section for more results, in particular for the multidimensional multiplicity ratio.} and $p_t$-broadening. The (one-dimensional) multiplicity ratio alone does not allow to put strong constraints on the production length $Lp$, and consequently on the amount on IEL. However, more stringent constraints are obtained after including HERMES data for the $p_t$-broadening. In the next section, we will see that the HERMES multidimensional multiplicity ratio also confirms our estimation of the IEL and the production length. At HERMES energies and $z=0.55$, the contribution of IEL to the hadron suppression is of the order of $7\%$ for heavy nuclei and $25\%$ for light nuclei. This contribution goes to 0 at $z=1$ due to energy conservation.

\section{More results: Kaons and 2 dimensional multiplicity ratios}\label{2dmultrat}
\begin{figure}[!h]
\centering
\includegraphics[width=8.3cm]{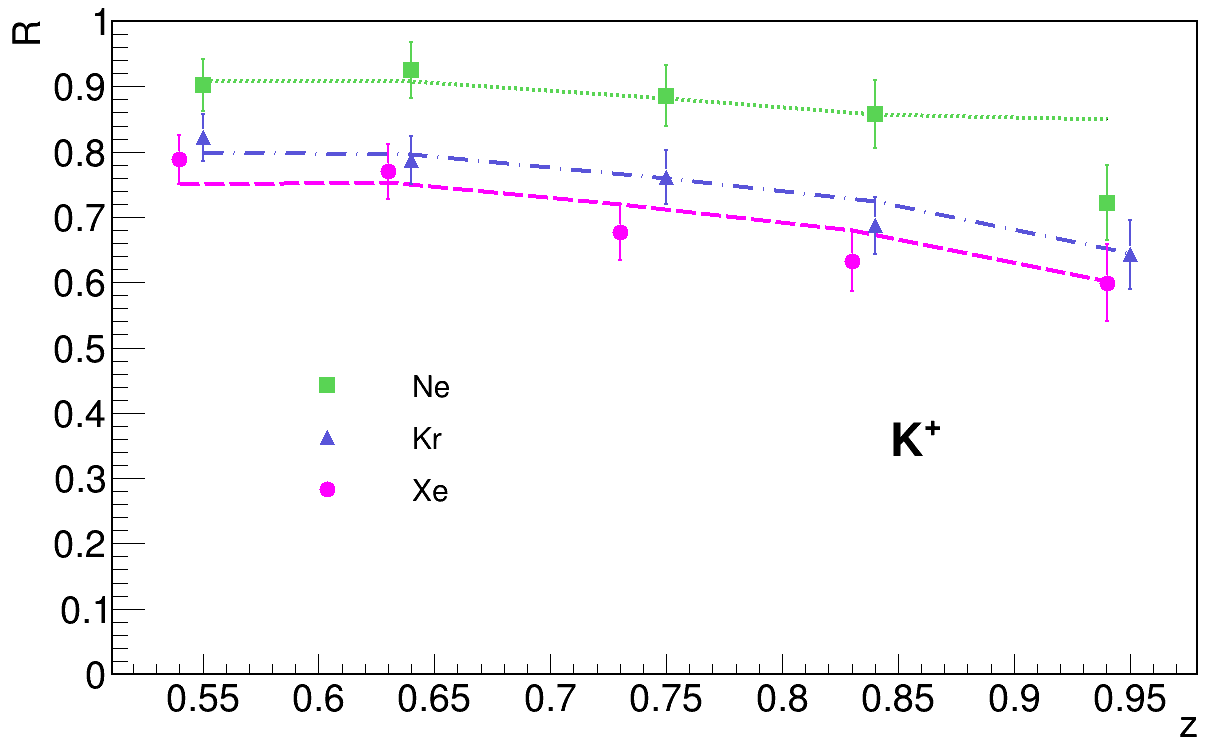}
\includegraphics[width=8.3cm]{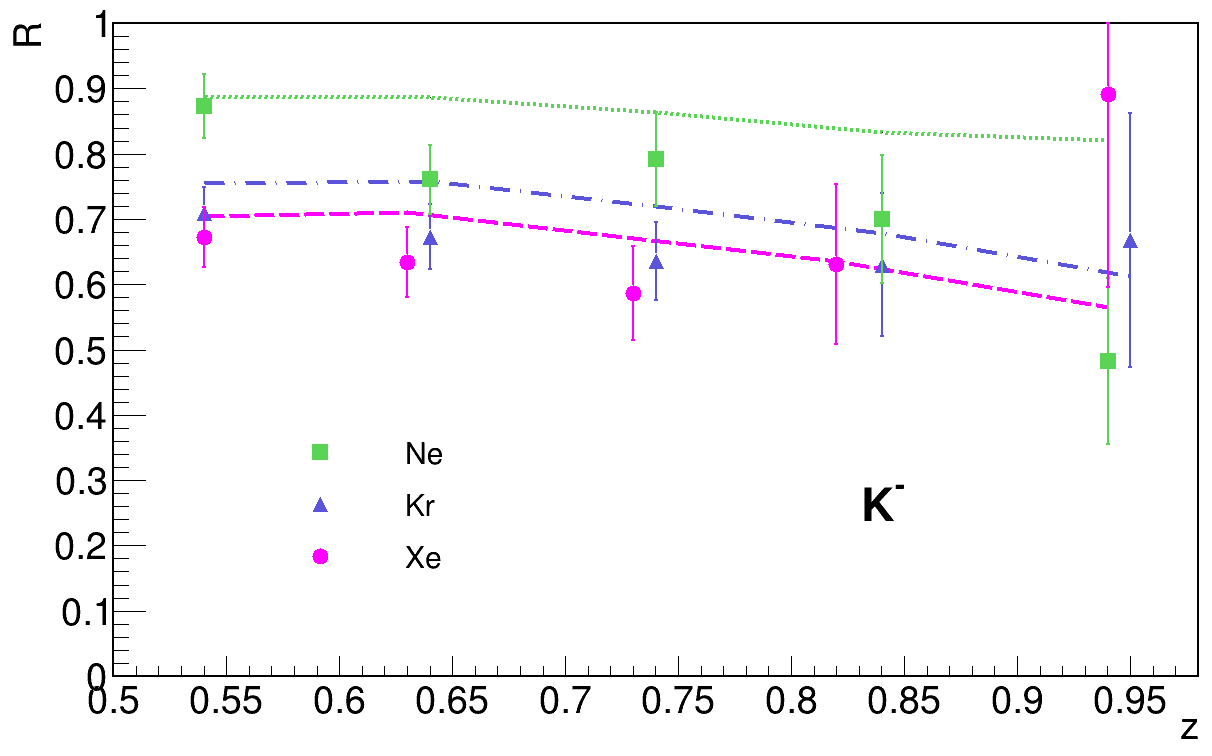}
\caption{HERMES multiplicity ratio for kaons.\label{hermkaon}}
\end{figure}

We start with HERMES data \cite{HER2} for $K^+$ and $K^-$, displayed in Fig. \ref{hermkaon} along with our calculations (using the same set of parameters).   
The error bars for the $K^-$ are larger due to the smaller cross section. It is due to the fact that this particle is made only of sea quarks, whose number densities are small at HERMES $x$ and $Q^2$. One of the interests of this observable is the stronger suppression of the $K^-$ compared to the $K^+$. While in IEL based models, this feature cannot be explained easily, it finds a very simple explanation in our case. The dipole inelastic cross section is proportional to the kaon-proton total cross sections. We used the parametrization given in \cite{pdg}, and the larger cross section for the $K^-$ gives the larger suppression seen in Fig. \ref{hermkaon}.\\

\begin{figure}[t!]
\hspace{-1.5cm}
\includegraphics[width=20cm]{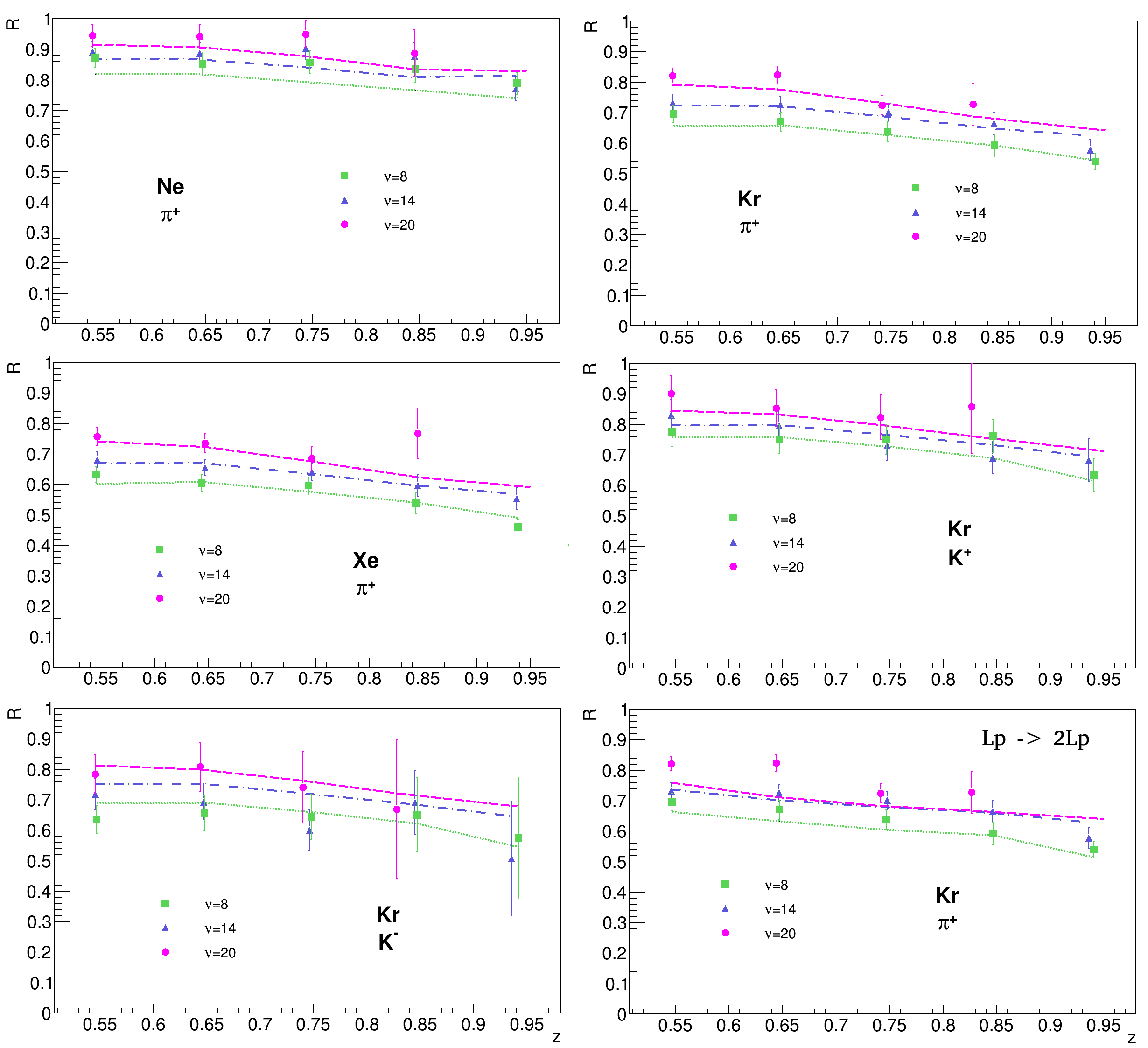}
\caption{$z$-dependence of the multiplicity ratio for different energy slices. The bottom right panel shows the same data as the top right panel, but with our calculations based on a larger $Lp$.\label{her2d}}
\end{figure}

Next, in Fig. \ref{her2d} we present our results for the multidimensional multiplicity ratio \cite{HER3} as a function of $z$.\footnote{The data are available on the HERMES website. We will not show all the plots for a matter of space.} The $z$ dependence has been measured for three slices of $\nu$, [4,12], [12,17] and [17,23.5]. In our simulations we used $\nu=8, 14, 20$.

The results show the same hierarchy than the experimental data, and a satisfying quantitative agreement. The reasons for this hierarchy in energy are simple. First, the production length increases with $\nu$, making the path for the dipole inside the nucleus smaller. Consequently, the effect of nuclear absorption is reduced. Second, as shown in Eq. (\ref{af}), the expansion of the dipole transverse size is slower (at asymptotic energy is it frozen), giving a smaller absorption cross section, and reducing further the nuclear absorption contribution.\\

\begin{figure}[!h]
\centering
\includegraphics[width=10cm]{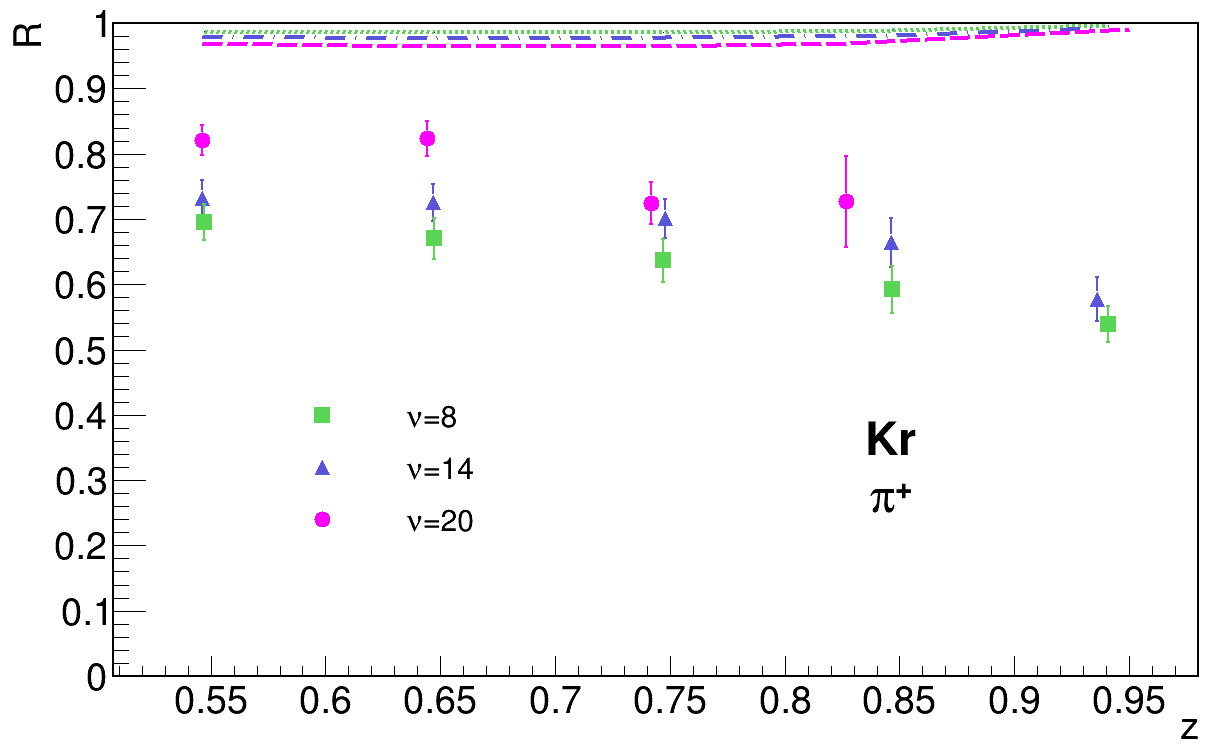}
\caption{Induced energy loss contribution to the HERMES 2-dimensional multiplicity ratio.\label{her2diel}}
\end{figure}

In our model, the IEL contribution alone gives a hierarchy contrary to the one seen in data, as demonstrated in Fig. \ref{her2diel}.
This behaviour is natural since, in our case, the path of the quark inside the nuclear medium increases with the energy, giving a increasing IEL contribution. However, our prediction for the IEL, Fig. \ref{her2diel}, is opposite to the one made by IEL based models, which predict that the suppression due to IEL decreases with energy. In these models, the quark travels (nearly\footnote{Some papers, e.g. \cite{Arl}, consider a "realistic" production length, which can be smaller than $L_{max}$, where $L_{max}$ corresponds to the travel throughout the whole nucleus. However in practice, for $\nu \gtrsim 10$ GeV $Lp=L_{max}$.}) throughout the whole nucleus. Then, increasing the energy does not increase the induced energy loss.\footnote{The quenching weight, appearing in Eq. (\ref{ielmod}), is small for large $\epsilon$, making the upper bound of the integral irrelevant.} Consequently, the shift in $z$
\begin{equation}
\Delta z \simeq z\frac{\Delta E_{ind}}{E},
\end{equation}
becomes smaller at higher energy, leading to a smaller suppression. However, we have seen that such large production lengths are disfavored by $p_t$-broadening data. By playing again the game of multiplying our $Lp$ by 2, we increase the contribution of IEL (which has the wrong hierarchy), and the agreement with data is lost, see Fig. \ref{her2d} (bottom right). This gives another confirmation that our estimation of the production length looks correct.\\

Finally, note that the absorption cross section has a very small energy dependence since $\sigma_{tot}^{\pi^+p}(\nu=8)>\sigma_{tot}^{\pi^+p}(14)>\sigma_{tot}^{\pi^+p}(20)$, $\sigma_{tot}^{K^-p}(8)>\sigma_{tot}^{K^-p}(14)>\sigma_{tot}^{K^-p}(20)$, but $\sigma_{tot}^{K^+p}(8)<\sigma_{tot}^{K^+p}(14)<\sigma_{tot}^{K^+p}(20)$, the last set of inequalities going in the opposite direction of the hierarchy observed in data. It could explain why the distance between the 3 lines at $z=0.55$ for the $K^+$ is smaller in comparison to the $\pi^+$ and $K^-$ particles.

\section{Consequences for the LHC\label{lhcphy}}
One of the interests of this paper, is to motivate the study of nuclear absorption at the LHC by a larger community. We believe that, if this effect is generally considered to be negligible, it is probably due to this kind of reasoning: 1) Nuclear absorption decreases with energy and 2) it is already small at medium energies. However, the second statement is not in agreement with the results presented in this paper.

To exemplify our claim, consider for instance \cite{Wan2}. In this paper, the main topic is the jet quenching at RHIC due to IEL. However, a plot for HERMES is shown in order to validate the choice of a very large $Lp$ (making nuclear absorption negligible). Here, we can see that the study of jet quenching at high energies relies partially on hypothesis at lower energies. We already discussed that fact that such large values for $Lp$ are in contradiction with HERMES data.\\

In fact, the whole reasoning could be incorrect, because the kinematics, the observables or the hypothesis used at the LHC can be quite different from the DIS case. One example is the jet production at central rapidity, $y=0$. If we write $p_t$, the parton transverse momentum, and $Q^2$ its virtuality, then we have the relation $E^2/2=p_t^2=Q^2$.\footnote{The last equality, $Q^2=p_t^2$, is not a strict one, imposed by 4-momentum conservation. But it is used in realistic Monte-Carlo simulations in order to reproduce the correct parton multiplicity in a jet.} We see that $E$ and $Q^2$ are correlated, while they can vary independently (up to certain limits) in the DIS case. This correlation has an important consequence, since $Lp$ increases/decreases for increasing $E$/$Q$. For large values of the virtuality, the vacuum energy loss could be so large that $Lp$ stays short, see \cite{KoNePoSc} for a more detailed discussion. The fact that the radiation process stops very quickly is confirmed by Monte-Carlo studies implementing explicitly parton branchings. It is the case in \cite{RoGoGoAi}, where the authors found that for $E=Q=100$ GeV, the cascade stops after a length of 1-3 fm.\\

Another interesting case is the quarkonia suppression in pA collisions at the LHC. First, we note that what is called nuclear absorption in some papers is not the same nuclear absoprtion adressed in studies on SIDIS experiments. For instance, in \cite{Fer,ArPe1}, what is called nuclear absorption is the absorption of the hadron by the medium. The hadron formation length being large for leading quarks, $l_f\propto zE$, the hadron is formed outside of the nucleus and cannot be absorbed. In the opposite, in studies of SIDIS experiments, the absoprtion concerns the hadron and the prehadron (dipole) \cite{GaMo,AcGrMu,PiGr}. The difference is that the latter can have a short production length $Lp$, even at high energy, due to the factor $(1-z)$ in Eq. (\ref{Lpz}). The absorption of the prehadron, the main effect at HERMES, is generally not considered at the LHC. It is for instance the case in \cite{ArPe1}, where the authors discuss the $J/\psi$ absorption in pA collisions. In this model, a color octet $c$-$\bar{c}$ pair propagates through the whole nucleus and the $J/\psi$ is formed outside. The absorption of the color-octet dipole by the medium is not discussed and the nuclear absorption is said to be negligible.

Considering the conclusion made by IEL based models for the multiplicity ratio at low/moderate energies, i.e the absorption of the prehadron is negligible\footnote{Generally, the statement is that the data are reproduced with induced energy loss. It implies that the nuclear absorption does not contribute.}, models like \cite{ArPe1} do not look inconsistent. The interest of our study is to show that the nuclear absorption is the main physical effect at HERMES energies. It implies that the discussion on the absorption of the color-octet dipole cannot be ignored at the LHC.\footnote{One could invoke a very small transverse size dipole, giving a negligible dipole cross section. However, it is generally accepted (see for instance \cite{Dok2}) that $|\vec{p_t}|\propto 1/r_t$, with $r_t$ the transverse size and $|\vec{p_t}|$ the transverse momentum of the quark in the $c$-$\bar{c}$ restframe. Then a very small dipole implies a very large mass.} It has been discussed in details in \cite{KoScSi}.\\

Finally, we note that at the workshop Hard Probes 2018, experimental results \cite{hardprobes} have been presented, showing a similar suppression in AA collisions for D meson and $J/\psi$ particles. The given explanation is that the wave function of the prehadron does not play any role, the suppression being a pure IEL effect. But then, the different $R_{AA}$ for the $J/\psi$ and the $\psi(2S)$ has been interpreted as a manifestation of the different wave functions. This inconsistency is not present in models considering nuclear absorption of color singlet/octet dipoles. It gives another motivation for calculations including this effect at the LHC.

\section{Predictions for the future EIC experiment}\label{eic}

\begin{figure}[!h]
\centering
\includegraphics[width=10cm,height=6cm]{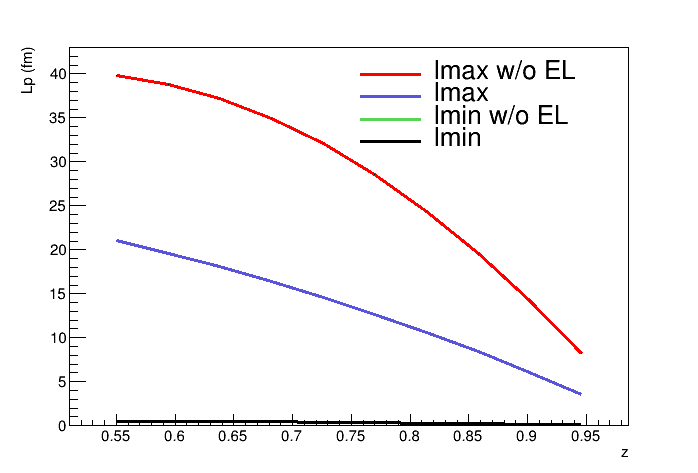}
\caption{Results for minimum and maximum values of production length, in fm, as a function of $z$. 
The green line is hidden by the black line. The kinematic is $Q^2=40$ GeV$^2$ and $\nu=100$ GeV, and the calculations are for lead. The red line has been obtained in the Born approximation (no energy loss).}\label{lpeic}
\end{figure}

The future electron ion collider will have a large kinematical range with $Q^2$ from 8 to 45 and $\nu$ from 30 to 150 GeV. In figure \ref{lpeic}, we present our results for the $z$ dependence of $Lp_{min}$ and $Lp_{max}$ for a lead nucleus, choosing $Q^2=40$ GeV$^2$ and $\nu=100$ GeV. Compared to HERMES, $Lp_{max}$ is several times larger. However, $Lp_{min}$ is still small enough to allow the production of the pre-hadron inside the nucleus.
We observe that the effect of energy loss are more important compared to HERMES. It is probably due to the fact that a larger $Q^2$ induces more vacuum energy loss.\\

\begin{figure}[!h]
\centering
\includegraphics[width=8cm,height=6cm]{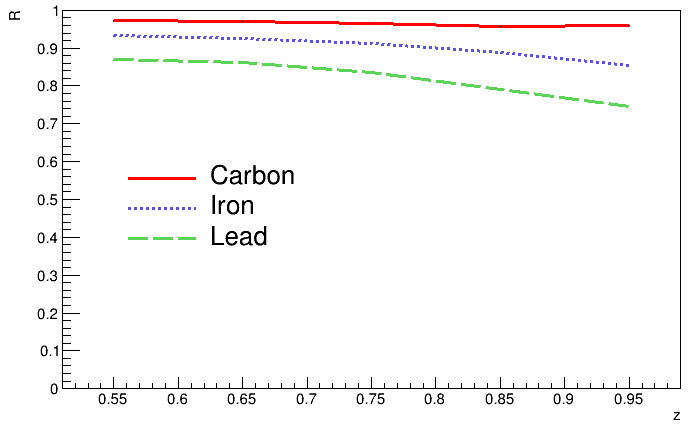}
\includegraphics[width=8cm,height=6cm]{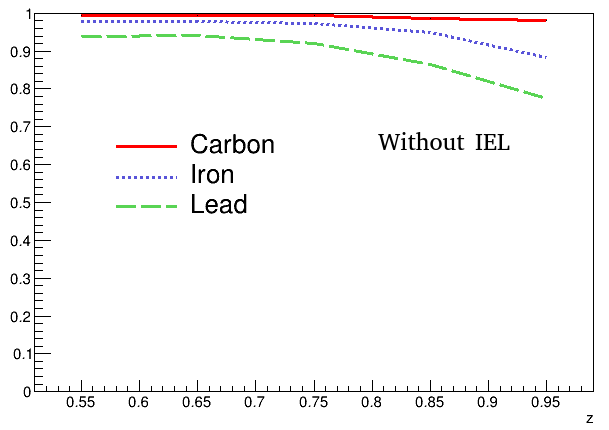}
\caption{Results for the multiplicity ratio for $Q^2=40$ GeV$^2$ and $\nu=100$ GeV. The right plot has been computed without IEL.}\label{reic}
\end{figure}

Figure \ref{reic} shows our predictions for the multiplicity ratio for $Q^2=40$ GeV$^2$ and $\nu=100$ GeV, with and without induced energy loss.
We can see that in this case, at $z=0.55$, the contribution of induced energy loss is approximately 50$\%$ for lead. As before the effect of IEL is more important for light nuclei. The increased contribution of IEL at the EIC was of course expected due to the increase of $Lp$ with energy. However this effect does not compensate the decrease of nuclear absorption, and the multiplicity ratio is larger (less suppression) than the one at CLAS. One of the interests of the EIC is that the kinematics is compatible with the LHC.\\

\begin{figure}[!h]
\centering
\includegraphics[width=9cm]{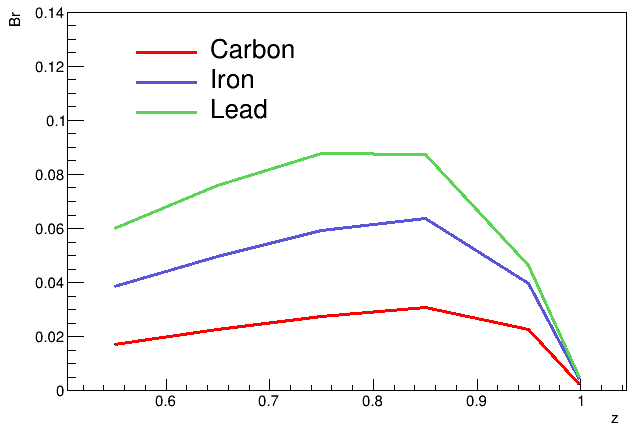}
\caption{Quark $p_t$-broadening at $Q^2=40$ GeV$^2$ and $\nu=100$ GeV.}\label{eicBr}
\end{figure}

In figure \ref{eicBr} we show the result for quark $p_t$-broadening. 
We expect this result to give an approximate estimation of the $p_t$-broadening which will be measured at the EIC. However, for quantitative description of this observable, one should go beyond than the simple quark $p_t$-broadening (for instance, the transverse kick due to hadronization is not taken into account). Note that our formalism includes the rise of $\hat{q}$ (related to the dipole cross section) with energy.

%\newpage

\section{Conclusion}
We have presented a model, including both induced energy loss and nuclear absorption, able to reproduce HERMES data for the multiplicity ratio, the 2-dimensional multiplicity ratio and the $p_t$-broadening. The model contains two free parameters, fixed at the same values for all observables. We studied the dependence of our results on these parameters, figure \ref{herm2}, and we observed a weak dependence on $x_0$.\\

The main goal of this study was the quantitative study of the nuclear absorption and induced energy loss contributions to the multiplicity ratio. It is closely related to the determination of the production length , $Lp$, of a colorless prehadron. The conclusion is:
\begin{itemize}
\item The relative contribution of these two effects depends on the nuclear size and on the kinematics, in particular on the beam energy and on the energy fraction $z$.
\item At HERMES energies, we found that $Lp \sim 2$ fm, and that the dominant effect to the multiplicity ratio is the nuclear absorption. The contribution of induced energy loss is approximately 7\% and 25\%, at $z=0.55$, for heavy and light nuclei, respectively. At $z\rightarrow 1$, $Lp$ goes to 0 and the IEL does not contribute.
\item Our conclusion does not agree with the one obtained by IEL based models, showing that induced energy loss is the main effect. This is due to the very large production length used in these calculations. We have shown that such large $Lp$ is not supported by HERMES data, in particular because it gives a too large contribution to the $p_t$-broadening. We also mentioned that studies like \cite{RoGoGoAi}, taking into account the vacuum energy loss, supports small production lengths. 
\item At EIC energies, the IEL can be the main contribution to the multiplicity ratio, in particular for light nuclei (except when $z\rightarrow 1$).
\item This does not implies that the IEL is always the main contribution at the LHC. We mentioned that, e.g due to different kinematical configurations, the nuclear absorption could still plays an important role.
\end{itemize}
In section \ref{2dmultrat}, we have presented our results for the 2-dimensional multiplicity ratio, as a function of $z$ and for different slices of $\nu$. The agreement with data is very satisfying, even for kaons. The increase of the multiplicity ratio with energy is interpreted as follow. At HERMES energies, the main effect responsible for the hadron suppression is the nuclear absorption. This effect decreases with $Lp$ and then with the energy. The interesting observation is that, in our model, and contrary to IEL based models, the energy dependence of the IEL contribution is opposite to the one observed in data (but the "IEL + nuclear absorption" contribution does have the correct energy dependence). The consequence is that, additionally to the $p_t$-broadening, the 2-dimensional multiplicity ratio allows to put more constraints on $Lp$. Indeed, a larger production length gives more induced energy loss, breaking the agreement with data due to its opposite behavior with energy, as shown in the bottom right panel of Fig. \ref{her2d}.

\section*{Acknowledgment}
We would like to thank Will Brooks for bringing to our knowledge the existence of the HERMES data for the 2-dimensional multiplicity ratio. We are also grateful to Hayk Hakobyan and Ahmed El Alaoui for sharing their knowledge of SIDIS experiments with us. We would like to thank Klaus Werner for his comments on the manuscript.
B. Guiot acknowledges support from Chilean FONDECYT iniciacion grants 11181126. B. Guiot acknowledges support by the Basal project FB0821.  B. Kopeliovich is  partially  supported  by  Fondecyt  grant  No. 1170319 and by Proyecto Basal FB 0821.

%\section*{References}

\end{document}